\documentclass[twocolumn,showpacs,preprintnumbers,amsmath,amssymb]{revtex4}

\usepackage{verbatim}

\newcommand{\half}{\frac{1}{2}}

\newcommand{\gap}{\vspace{3mm}}

\newcommand{\odo}[1]{\ensuremath{\frac{d}{d #1}}}
\newcommand{\odho}[2]{\ensuremath{\frac{d^{#1}}{d #2^{#1} }}}

\newcommand{\pdo}[1]{\ensuremath{\frac{\partial }
        {\partial #1 }}}

\newcommand{\slashletter}[1]{\ensuremath{\kern+0.1em /\kern-0.65em #1}}

\begin{document}

\title{Fermion determinant for general background gauge fields}

\author{M. P. Fry}

\affiliation{School of Mathematics, University of Dublin, Dublin 2, Ireland}

\date{\today}

\begin{abstract}
An exact representation of the Euclidean fermion determinant in two
dimensions for centrally symmetric, finite-ranged Abelian background
fields is derived. Input data are the wave function inside the field's
range and the scattering phase shift with their momenta rotated to the
positive imaginary axis and fixed at the fermion mass for each
partial-wave. The determinant's asymptotic limit for strong coupling
and small fermion mass for square-integrable, unidirecitonal magnetic
fields is shown to depend only on the chiral anomaly. The concept of
duality is extended from one to two-variable fields, thereby relating
the two-dimensional Euclidean determinant for a class of background
magnetic fields to the pair production probability in four dimensions
for a related class of electric pulses. Additionally, the
``diamagnetic'' bound on the two-dimensional Euclidean determinant is
related to the negative sign of
$\partial \text{Im}S_{\text{eff}}/\partial m^2$ in four dimensions in
the strong coupling, small mass limit, where $S_{\text{eff}}$ is the
one-loop effective action.
\end{abstract}


\maketitle

\section{\label{Sec_SecI}
Introduction}

Within the Standard Model fermion determinants are encountered in
the calculation of every physical process. Because of their nonlocal
dependence on the gauge fields they are difficult to calculate.
Consequently the practice has been to either ignore them--the quenched
approximation--or to expand them in power series. Ultimately they will
have to be confronted nonperturbatively, as lattice theorists are now
doing with faster machines, in order to obtain reliable predictions
with known computational error.

The current status of fermion determinants is reviewed
in~\cite{Fry02}. To indicate just how bad our knowledge of these
determinants is, not even the strong coupling limit of the massive
Euclidean QED determinant in two dimensions is known except for a
constant background magnetic field and for a magnetic field confined
to the surface of a cylinder~\cite{Fry95}.  Therefore, it seems that
this is as good a starting point as any to get a better insight into
the properties of fermion determinants and to understand their
physics.

There are other reasons why the $\text{QED}_{2}$ determinant should be
of general interest. Namely, if there were precise nonperturbative
information on at least one continuum, infinite-volume determinant
then the algorithms of lattice theorists for calculating determinants
could be tested by extrapolating their output to zero lattice spacing
and infinite volume. Algorithms for determinants can be easily
adjusted to any dimensionality, and if some fail to coincide with
known results for an Abelian background field in two dimensions then
they are certainly useless.

Work in this direction has already begun~\cite{Chiu00} with the
computation of the fermion determinant for massless fermions on a
torus using the Neuberger-Dirac operator and the higher-order overlap
Dirac operator and the comparison of the results with the exact
massless QED$_2$ determinant on a torus~\cite{Sachs92}. In massive
two-flavour QED$_2$ the determinant was calculated explicitly to study
the masses of the triplet (pion) and singlet (eta) bound states
using the overlap and fixed point Dirac
operators~\cite{Hausler01}. Presumably the continuum limit of the
determinant itself in the nonperturbative domain discussed below could
be used as a sensitive test of the many lattice discretizations of the
Dirac operator now in use.

In addition, we develop further the concept of duality and relate the
Euclidean $\text{QED}_{2}$ determinant to the pair production
probability in $\text{QED}_{4}$ for a class of electric pulses. Thus,
the Euclidean $\text{QED}_{2}$ determinant contains nonperturbative
physical information in four dimensions.

Fermion determinants are obtained by integrating over the fermion
fields to produce the one-loop effective Euclidean action
$S_{\text{eff}} = -lndet$, where $det$ is formally the ratio
$det(\slashletter{P} - e\slashletter{A} + m)/det(\slashletter{P} + m)$
of Fredholm determinants of Euclidean Dirac operators. We assume that
the continuation to the Euclidean metric has been done. When $det$ is
properly defined it is a nonlocal function of the field strength
$F_{\mu\nu}$ formed from the potential $A_{\mu}$ , modulo Chern-Simons
terms that are absent in two dimensions. Since the determinant is part
of the gauge field's action, $A_{\mu}$ and $F_{\mu\nu}$ are random
fields. We have discussed elsewhere~\cite{Fry95,Fry92,Fry96} how the
need to regulate in any dimension above one allows one to assume
smooth potentials and fields. In order to make further progress we
assume in addition that $F_{\mu\nu}$ is centrally symmetric and that
it has a finite range $a$.

This paper is organized as follows. In Sec. \ref{Sec_SecII} we define
the determinant and indicate our strategy for calculating it by first
assuming $ma<<1$ and then letting $|e \Phi| >> 1$, where $m$ is the
fermion mass and $\Phi$ is the flux of the background magnetic field
$F_{12}$. This is the really interesting limit as it takes one deep
into the nonperturbative regime. In Sec. \ref{Sec_SecIII} the
low-energy scattering phase shifts required to calculate the
determinant are obtained. Section \ref{Sec_SecIV} deals with the small
mass, strong coupling expansion of the determinant, while
Sec. \ref{Sec_SecV} presents the explicit form it takes in this
limit. Section \ref{Sec_SecVI} generalizes the concept of duality from
one to two-variable fields, thereby allowing the QED$_2$ Euclidean
determinant to be related to physics in four dimensions. Section
\ref{Sec_SecVII} summarizes our results while the asymptotic form of
the determinant given in Sec. \ref{Sec_SecV} is derived in the
Appendix.

\section{\label{Sec_SecII}
Representation of the Determinant}

\subsection{\label{SubSec_GreenFunc}
Green's functions}

The exact calculation of $det$ in $\text{QED}_2$ continued to the
Euclidean metric reduces to the scattering problem of a charged
particle confined to a plane pierced by a magnetic field,
namely~\cite{Fry93}

\begin{widetext}
\begin{equation}
\label{Eq_0201}
\pdo{e} lndet =
  \frac{e}{\pi}
  \int \, d^2r \, \varphi \, \partial^2 \varphi +
  2 m^2 \int \, d^2r \, \varphi(\textbf{r})\\
  <\textbf{r} | (H_+ + m^2)^{-1} - (H_- + m^2)^{-1}|\textbf{r}>,
\end{equation}
\end{widetext}

\noindent
where the supersymmetric operator pair
$\text{H}_{\pm} = (\textbf{P} - e\textbf{A})^2 \mp eB$ are obtained
from the two-dimensional Pauli Hamiltonian
$(\textbf{P} - e\textbf{A})^2 - \sigma_3 eB$. Hence, the
subscripts on H in (\ref{Eq_0201}) refer to positive and negative
chirality. The auxiliary potential $\varphi$ is related to the vector
potential by $A_{\mu} = \epsilon_{\mu\nu} \partial_\nu \varphi$
and to the magnetic field by $B = - \partial^2 \varphi$ or

\begin{equation}
\label{Eq_0202}
\varphi(\textbf{r}) =
	-\frac{1}{2\pi} \int d^2r'
	\ln |\textbf{r} - \textbf{r}'| B(\textbf{r}'),
\end{equation}

\noindent
with $\epsilon_{12} = 1$. Expansion of (\ref{Eq_0201}) in powers of $e$
yields the standard one-loop effective action given by the Feynman
rules. The first term on the right-hand side of (\ref{Eq_0201}) is
$\partial lndet / \partial e$ of the massless Schwinger
model~\cite{Schwinger62}. Due to the $1/r$ falloff of $A_{\mu}$ when
$\Phi \ne 0$ an integration by parts is not justified in this case. As
we will see in Sec. \ref{Sec_SecV}, the presence of the mass dependent
term profoundly modifies the determinant, ultimatelly cancelling the
first term when $|e \Phi| >> 1$. The invariance of (\ref{Eq_0201})
under $\varphi \to \varphi + c$, where $c$ is a constant, gives the
index theorem on a two-dimensional Euclidean manifold~\cite{Fry93,Musto86}.

We now assume that $B$ is centrally symmetric and that $B(r) = 0$ for
$r > a$. To ensure finite flux we assume $B$ is square-integrable in view
of the inequality
$\Phi^2 \leq 2 \pi^2 a^2 \int^a_0 \, dr \, r \, B^2(r)$.
Referring to (\ref{Eq_0201}), define the Green's function

\begin{eqnarray}
\nonumber
\lefteqn{<r,\theta |(k^2 - H_{\pm})^{-1}| r',\theta'>}        \\
\nonumber
	&=& \frac{1}{2 \pi} \sum^{\infty}_{l=-\infty}
	<r | (k^2 - H_{\pm,l})^{-1} | r'> e^{il(\theta - \theta')}\\
\label{Eq_0203}
	&=& \frac{1}{2 \pi} \sum^{\infty}_{l=-\infty}
            G_{\pm,l}(k;r,r') e^{il(\theta - \theta')},
\end{eqnarray}

\noindent
where $A_{\theta} = \Phi(r) / 2 \pi r,$

\begin{equation}
\label{Eq_0204}
H_{\pm, l} = - \odho{2}{r} - \frac{1}{r} \odo{r}
	+ \frac{(l - e\Phi(r) / 2\pi)^2}{r^2} \mp e B(r),
\end{equation} 

and

\begin{equation}
\label{Eq_0205}
\Phi(r) = 2 \pi \int^{r}_{0} \, ds \, s \, B(s). 
\end{equation}

The calculation is simplified by introducing the Green's function

\begin{equation}
\label{Eq_0206}
\mathcal{G}_{\pm,l}(k;r,r') = \sqrt{rr'} G_{\pm,l}(k;r,r'),
\end{equation}

\noindent
where

\begin{equation}
\label{Eq_0207}
\mathcal{G}(k;r,r') = <r|(k^2 - \mathcal{H}_{\pm,l})^{-1}|r'>,
\end{equation}

\noindent
and

\begin{equation}
\label{Eq_0208}
\mathcal{H}_{\pm,l} = - \odho{2}{r}
	+ \frac{(l - e \, \Phi(r) / 2\pi)^2 - \frac{1}{4}}{r^2} \mp e B(r).
\end{equation}

The outgoing-wave Green's functions $\mathcal{G}_{\pm, l}$ are
constructed from ~\cite{Newton82}

\begin{equation}
\label{Eq_0209}
\mathcal{G}_{\pm,l}(k;r,r')
	= - \frac{\varphi(k, r_{<}) f^{(+)}(k, r_{>})}{\mathcal{J}(k)},
\end{equation}

\noindent
where $\varphi$ is a regular solution and $f^{(+)}$ an irregular
outgoing-wave solution, of

\begin{equation}
\label{Eq_0210}
\mathcal{H}_{\pm,l} f = k^2 f;
\end{equation}

\noindent
$\mathcal{J}$ is the associated Jost function and $r_{<}$, $r_{>}$ denote the lesser and
larger values of $r, r'$. Here and below we will occasionally suppress
the subscripts $\pm$ and $l$ to reduce notational clutter.

Regular solutions of (\ref{Eq_0210}) are

\begin{widetext}
\begin{equation}
\label{Eq_0211}
\varphi_{\pm, l}(k,r) = 
\frac{e^{i\pi(|l| - W/2)}}{2\sqrt{2}}
\begin{cases}
\displaystyle
\sqrt{a} \left( H^{-}_{W} (ka) + S_{\pm} H^{+}_{W} (ka) \right)
  \frac{R_{\pm}(k,r)}{R_{\pm} (k,a)}, & r < a	\\[3mm]
\displaystyle
\sqrt{r} \left( H^{-}_{W} (kr) + S_{\pm} H^{+}_{W} (kr) \right), 
  & r > a,
\end{cases}
\end{equation}
\end{widetext}

\noindent
where

\begin{equation}
\label{Eq_0212}
S_{\pm} = e^{i\pi(W - |l|)} e^{2 i \delta^{\pm}_{l}};
\end{equation}

\noindent
$H^{+}_W$ and $H^{-}_W$ denote the Hankel functions $H^{(1)}_{W}$
and $H^{(2)}_{W}$, respectively; $\delta^{\pm}_{l}$ are the
scattering phase shifts; $W = |l - e \Phi / 2 \pi|$, and
$\Phi = \Phi(a)$ is the total flux of $B$. The interior wavefunctions
$R_{\pm,l}$ satisfy the boundary condition
$\lim_{r \to 0} r^{-\half -|l|} R_{\pm, l} = 1$. These will be
discussed further below. The structure of $\varphi_{\pm,l}$ for
$r > a$ ensures that the eigenfunctions of $H_{\pm, l}$ in
(\ref{Eq_0204}), $\psi_{\pm,l} = \varphi_{\pm,l} / \sqrt{r}$,
correspond to physical wave functions~\cite{Newton82}. That is,

\begin{equation}
\label{Eq_0213}
\psi_{\pm}(\mathbf{k}, \mathbf{r})
  = \frac{1}{\sqrt{2}\pi} \sum^{\infty}_{l=-\infty}
  \psi_{\pm,l}(k,r) e^{il \theta},
\end{equation}

\noindent
assumes the asymptotic form for $r \to \infty$

\begin{equation}
\label{Eq_0214}
\psi_{\pm}(\mathbf{k}, \mathbf{r})
  \sim \frac{1}{2\pi} e^{i\mathbf{k}.\mathbf{r}}
  + \frac{1}{2\pi\sqrt{r}} f_{\pm}(k, \theta) e^{ikr},
\end{equation}

\noindent
where

\begin{equation}
\label{Eq_0215}
f_{\pm}(k, \theta)
  = \sqrt{\frac{2}{\pi k}} e^{\frac{i\pi}{4}}
  \sum^{\infty}_{l = -\infty} e^{i\delta^{\pm}_{l}}
  \text{sin} \, \delta^{\pm}_{l} e^{il \theta}, 
\end{equation}

\noindent
with the differential scattering cross section
$d\sigma/d\Omega = |f_{\pm}(k,\theta)|^2$.

Assuming $R_{\pm, l}$ are known, irregular outgoing-wave solutions of (\ref{Eq_0210})
can be found by standard means, giving

\begin{widetext}
\begin{equation}
\label{Eq_0216}
f^{(+)}_{\pm, l} (k, r) =
\begin{cases}
\displaystyle
\sqrt{a} \, H^{+}_{W}(ka) \, \frac{R_{\pm}(k,r)}{R_{\pm}(k,a)} 
  + \frac{4i}{\pi\sqrt{a}}
  (H^{-}_{W}(ka) + S_{\pm} H^{+}_W(ka))^{-1}
  R_{\pm}(k,a) R_{\pm}(k,r)
  \int^a_r \frac{ds}{R^2_{\pm}(k,s)}, & r < a,\\[3mm]
\sqrt{r} \, H^{+}_{W}(kr), & r > a.
\end{cases}
\end{equation}
\end{widetext}

\noindent
Near the regular singular point at $r = 0$ of (\ref{Eq_0210}),
$f^{(+)}_{\pm,l} \sim const \; \times \; r^{\half - |l|}$.

Equations (\ref{Eq_0211}) and (\ref{Eq_0216}) give the Jost function

\begin{equation}
\label{Eq_0217}
\mathcal{J} = W(f^{(+)}, \varphi)
	= - \frac{i\sqrt{2}}{\pi} e^{i\pi(|l| - W/2)},
\end{equation}

\noindent
which is independent of chirality; $W$ on the left-hand side is the
Wronskian. It may be verified that (\ref{Eq_0209}), (\ref{Eq_0211}),
(\ref{Eq_0216}) and (\ref{Eq_0217}) combine to satisfy the basic
condition

\begin{equation}
\label{Eq_0218}
\pdo{r} \mathcal{G}_{\pm,l}(r, r')
	|^{r = r' + 0}_{r = r' - 0} = 1.
\end{equation}

In order to make contact with the determinant in (\ref{Eq_0201}) we now
analytically continue $k$ in $\mathcal{G}_{\pm,l}(k,r)$ into the upper
half of the complex plane by letting $k = me^{i\pi/2}$. Then
(\ref{Eq_0201}), (\ref{Eq_0203}) and (\ref{Eq_0206}) give

\begin{widetext}
\begin{equation}
\label{Eq_0219}
\pdo{e} lndet = -2e \int^{a}_{0} dr \, r \varphi(r) B(r)
  -2 m^2 \, \int^{\infty}_0 dr \, \varphi(r) \,
  \sum^{\infty}_{l=-\infty}
  \left[
  \mathcal{G}_{+,l} (me^{i\pi/2},r) - \mathcal{G}_{-,l} (me^{i\pi/2},r)
  \right],
\end{equation}
\end{widetext}

\noindent
while (\ref{Eq_0202}) gives

\begin{equation}
\label{Eq_0220}
\varphi(r) =
\begin{cases}
\displaystyle
\frac{1}{2\pi} \int^a_r ds \, \frac{\Phi(s)}{s},& r < a\\[3mm]
\displaystyle
- \frac{\Phi}{2\pi} \ln\left(\frac{r}{a}\right), & r > a.
\end{cases}
\end{equation}

\noindent
Because of the invariance of $lndet$ under $\varphi \to \varphi + c$
we have adjusted $\varphi(r)$ so that $\varphi(a) = 0$.

For $r > a$, (\ref{Eq_0209}), (\ref{Eq_0211}), (\ref{Eq_0216}) and
(\ref{Eq_0217}) give

\begin{widetext}
\begin{equation}
\label{Eq_0221}
\mathcal{G}_{+,l}(me^{i\pi/2}, r) - \mathcal{G}_{-,l}(me^{i\pi/2},r)
  = \frac{ir}{\pi} \, e^{-i\pi|l|}
  (e^{2i\delta^{+}_l} - e^{2i\delta^{-}_l}) \, K^2_W(mr),
\end{equation}
\end{widetext}

\noindent
where $K_W$ is a modified Bessel function and we used~\cite{Abramowitz64}

\begin{equation}
\label{Eq_0222}
H^{+}_W(rme^{i\pi/2}) = - \frac{2i}{\pi} e^{-i\pi W/2} K_W(mr).
\end{equation}

\noindent
The phase shifts in (\ref{Eq_0221}) are understood to be analytically
continued as well.

It is convenient to separate the energy-independent Aharonov-Bohm
phase shifts~\cite{Musto86,Jaroszewicz86} from
$\delta^{\pm}_l$. Without loss of generality we assume
$e \Phi > 0$. Then, modulo $\pi$,

\begin{equation}
\label{Eq_0223}
\delta^{+}_l(k) =
\begin{cases}
\displaystyle
\frac{\pi}{2}(|l| - W) + \Delta^{+}_l(k),& l \neq [e\Phi/2\pi]\\[5mm]
\displaystyle
\frac{\pi}{2}(e\Phi/2\pi) + \Delta^{+}_l(k),& l = [e\Phi/2\pi]\\
\end{cases}
\end{equation}

\gap
\gap

\begin{equation}
\label{Eq_0224}
\delta^{-}_l(k) = \frac{\pi}{2}(|l| - W) + \Delta^{-}_l(k),
	\hspace{1cm} \text{all } l,
\end{equation}

\noindent
where $[x]$ stands for the nearest integer less than $x$ with $[0] =
0$. The energy-dependent phase shifts $\Delta^{\pm}_l(k)$ will be
calculated in Sec. \ref{Sec_SecIII}.

The Green's function difference on the left-hand side of
(\ref{Eq_0221}) for $r < a$ may be dealt with as for $r > a$, this
time using ~\cite{Abramowitz64}

\begin{widetext}
\begin{equation}
\label{Eq_0225}
H^{+}_W(ame^{i\pi/2}) \, H^{-}_W(ame^{i\pi/2})
  = \frac{4}{\pi^2} e^{-i\pi W} K^2_W (am) - \frac{4i}{\pi} I_W(am) K_W(am), 
\end{equation}
\end{widetext}

\noindent
where $I_W$ is a modified Bessel function. For
$e\Phi/2\pi = N + \epsilon$, $N = 0, 1,...$; $0 \leq \epsilon < 1$
the final result from (\ref{Eq_0219}), (\ref{Eq_0220}),
(\ref{Eq_0221}), (\ref{Eq_0223}) and (\ref{Eq_0224}) is

\begin{widetext}
\begin{equation}
\label{Eq_0226}
\begin{split}
\pdo{e}lndet
  =& -2 e \int^a_0 dr \, r \, \varphi(r) B(r)
  + 2am^2 \int^a_0 dr \, \varphi(r) \sum_l I_W(am) K_W(am)
  \left[
  \left(\frac{R_{+}(r)}{R_{+}(a)}\right)^2 
  - \left(\frac{R_{-}(r)}{R_{-}(a)}\right)^2
  \right]\\
  &+ \frac{i2am^2}{\pi} \int^a_0 dr \, \varphi(r) \,
  \sum_{l \neq N} e^{-i \pi W} K^2_W(am)
  \left[
  (1 - e^{2i\Delta^{+}_l}) \left(\frac{R_{+}(r)}{R_{+}(a)}\right)^2
  - (1 - e^{2i\Delta^{-}_l}) \left(\frac{R_{-}(r)}{R_{-}(a)}\right)^2
  \right]\\
  &+ \frac{i2am^2}{\pi} e^{-i \pi \epsilon} K^2_{\epsilon}(am)
  \int^a_0 dr \, \varphi(r)
  \left[
  (1 - e^{2i\pi\epsilon} e^{2i\Delta^{+}_N})\left(\frac{R_{+}(r)}{R_{+}(a)}\right)^2
  - (1 - e^{2i\Delta^{-}_N})\left(\frac{R_{-}(r)}{R_{-}(a)}\right)^2
  \right]\\
  &+ 2 m^2 \int^a_0 dr \, \varphi(r)
  \sum_l
  \left[
  R^2_{+}(r) \int^a_r \frac{ds}{R^2_{+}(s)}
  - R^2_{-}(s) \int^a_r \frac{ds}{R^2_{-}(s)}
  \right]\\
  &+ \frac{i m^2 \Phi}{\pi^2} \int^{\infty}_a dr \, r \,
  \ln \, (r/a)
  \sum_{l \neq N}
  e^{-i\pi W}(e^{2i\Delta^{+}_l} - e^{2i \Delta^{-}_l})
  K^2_W(mr)\\
  &+ \frac{im^2 \Phi}{\pi^2} (e^{i\pi\epsilon}
  e^{2i\Delta^{+}_N} - e^{- i\pi\epsilon}e^{2i\Delta^{-}_N})
  \int^{\infty}_a dr \, r \,
  \ln \, (r/a) K^2_{\epsilon}(mr).
\end{split}
\end{equation}
\end{widetext}

\noindent
The interior wave functions $R_{\pm}$ and the phase shifts
$\Delta^{\pm}_l$ are abbreviations for $R_{\pm, l}(me^{i\pi/2}, r)$
and $\Delta^{\pm}_l(me^{i\pi/2})$.

The representation (\ref{Eq_0226}) is exact. Its advantage over other
representations of determinants based on scattering data is that it
involves no integration over phase shift energy. It is particularly
relevant to a study of the chiral limit $ma<<1$. Anticipating what follows,
the integrals can be interchanged with the sums for the class of fields
considered here, allowing the integrals in the exterior region $r>a$
to be done immediately. Only information about the interior wave
functions is required to calculate the determinant exactly, and these
are known explicitly for $ma<<1$ as in (\ref{Eq_0229}) below.

The right-hand side of (\ref{Eq_0226}) must be real since it is a
Euclidean determinant. This imposes the nontrivial constraints

\begin{widetext}
\begin{eqnarray}
\label{Eq_0227}
\nonumber
e^{i\pi W} e^{-2i \Delta^{\pm}_l(me^{-i\pi/2})}
  &=& -e^{-i\pi W} e^{2i \Delta^{\pm}_l(me^{i\pi/2})}
      + 2 \, \text{cos} \, \pi W, \hspace{5mm} l \neq N \,
      \text{for + chirality}\\
\nonumber
e^{-i\pi\epsilon} e^{-2i\Delta^{+}_N (me^{-i\pi/2})}
  &=& -e^{i\pi\epsilon} e^{2i\Delta^{+}_N (me^{i\pi/2})}
      + 2 \, \text{cos} \, \pi \epsilon,\\
\Delta^{\pm *}_l (me^{i\pi/2})
  &=& \Delta^{\pm}_l (me^{-i\pi/2}).
\end{eqnarray}
\end{widetext}

\noindent
For the fields considered here and the small mass expansions of
$\Delta^{\pm}_l$ made in Sec. \ref{Sec_SecIII} there is complete
agreement with (\ref{Eq_0227}).

\subsection{\label{SubSec_SmallMass}
Small mass expansions}

We now commence the expansion of $lndet$ when $ma<<1$. This does not
mean an expansion in powers of $m^2$. Such an expansion does not exist as
$lndet$ has a branch beginning at $m = 0$~\cite{Fry00a}. Rather, we are referring
to a collection of leading terms in $m$ such as $m^{\nu} \, \ln \, m$,
$\nu > 0$, as well as intergral powers of $m^2$.

Since (\ref{Eq_0210}) depends only on $k^2$ and the boundary condition
$\lim_{r \to 0} r^{-|l| - \half} R_{\pm,l} = 1$ is independent of
$k, R_{\pm,l}(k,r)$ is a regular function of $k^2$. Therefore we set
$R_{\pm,l}(me^{i\pi/2}, r) \equiv R_{\pm,l}(m^2,r)$ and begin an
expansion in powers of $m^2$:

\begin{equation}
\label{Eq_0228}
R_l(m^2,r) = R_l(r) (1 + (ma)^2 \chi_l(r) + O(ma)^4).
\end{equation}

\noindent
For $m = 0$, exact positive chirality solutions of
$\mathcal{H}_{+,l} R_{+,l} = 0$ are
known for $l > 0$~\cite{Jaroszewicz86}; the remaining cases can be
dealt with similarly. The results are, up to irrelevant normalization
constants that cancel in (\ref{Eq_0226}),

\begin{equation}
\begin{array}{llr}
\label{Eq_0229}
R_{+,l} &= r^{l + \half} \, e^{e \, \varphi(r)},& l \geq 0,\\[3mm]
\displaystyle
R_{+,-l} &= r^{-l + \half} \, e^{e \, \varphi(r)} \,
  \displaystyle
  \int^r_0 ds \; s^{2l-1} e^{-2e \, \varphi(s)},& l > 0,\\[3mm]
R_{-,l} &= r^{-l + \half} \, e^{-e \, \varphi(r)} \,
  \displaystyle
  \int^r_0 ds \; s^{2l-1} e^{2e \, \varphi(s)},& l > 0,\\[3mm]
R_{-,-l} &= r^{l + \half} \,  e^{-e \, \varphi(r)},& l \geq 0.
\end{array}
\end{equation}

\gap
\noindent
Noting (\ref{Eq_0220}), $R_{+,l}$ is square-integrable for
$l = 0, .., N-1$ for $e\Phi/2\pi = N + \epsilon$. This is in accord
with the Aharonov-Casher theorem which states that the number of
positive (negative) chirality square-integrable zero modes is
$[|e\Phi|/2\pi]$, depending on whether $e\Phi > 0$
$(e\Phi < 0)$~\cite{Aharonov79}. These zero modes will be shown to
play a dominant role in the strong coupling limit of $lndet$.

We want to calculate $lndet$ in the limit $ma<<1$ followed by
$e\Phi >> 1$. This must be done with care as there may be ratios of
terms like $(am)^2 e^{e\Phi} /(1 + (am)^2 e^{e\Phi})$ which when
further expanded in powers of $m^2$ grows exponentially with
$e\Phi$. There is one firm guiding principle here, namely that the
determinant is an entire function of $e$ of order
2~\cite{Seiler75,Seiler82}. This means that for any complex value of $e$,
$|\text{det}| < A(\epsilon) \, \text{exp}\left(K(\epsilon)|e|^{2 + \epsilon}\right)$
for any $\epsilon > 0$ and $A(\epsilon)$, $K(\epsilon)$ are
constants. Therefore, any growth of $lndet$ faster than quadratic in
$e$ means that the expansion one is making is inadmissible. In fact,
for real values of $e$ $lndet$ must satisfy the more precise bound

\begin{equation}
\label{Eq_0230}
-\frac{e^2||B||^2}{4\pi m^2} \leq lndet \leq 0,
\end{equation}

\noindent
for any $B$ with $||B||^2 = \int d^2r \, B^2(\mathbf{r}) < \infty$.
There are additional technical assumptions underlying
(\ref{Eq_0230}) that the fields considered here satisfy.
The right-hand side is the "diamagnetic"
bound~\cite{Seiler81,Seiler82,Brydges79,Weingarten80} and the
left-hand side follows from the general operator structure of $det$
and some standard inequalities~\cite{Fry02}.

The warning cited above materialzes for $0 < l < e\Phi/2\pi$
when $B(r) \geq 0$. There may be other cases. In the positive chirality
sector $\chi^{+}_l$ in (\ref{Eq_0228}) is

\begin{equation}
\label{Eq_0231}
\chi^{+}_l(r) = a^{-2} \int^r_0 ds \, \int^s_0 dt \,
  (t/s)^{2l + 1} e^{2e[\varphi(t) - \varphi(s)]},
\end{equation}

\noindent
for $r < a$. What happens is that the effective potential

\begin{equation}
\label{Eq_0232}
V(r) = \frac{(l - e\Phi(r)/2\pi)^2 - \frac{1}{4}}{r^2} - eB(r),
\end{equation}

\noindent
has a high and wide barrier beginning in the range $r < a$ and
extending out to $r \sim 2a$ for $e\Phi/2\pi >> 1$. This gives rise to
quasi-stationary states. As a consequence the wave function is
enhanced inside $r < a$ and $\chi^{+}_l$ can become large for strong
coupling.

For $l = O(e\Phi/2\pi)$ or larger the barrier in $V$ disappears and the
growth of $\chi^{+}_l$ for $e\Phi/2\pi >> 1$ slows down. This must
happen since $d\chi^{+}_l/dl < 0$ for all $e$. For $l >> 1$ the
integral in (\ref{Eq_0231}) is dominated in the range $t \lesssim s$, giving
$\chi^{+}_l = O(1/l)$. For the special case of $B(r) = B$, $r < a$ and
zero otherwise we find

\begin{equation}
\label{Eq_0233}
\chi^{+}_l(r) \leq (4l)^{-1} \ln \, l + O(1/l),
\end{equation}

\noindent
for $l > e\Phi/2\pi - 1$, $l > 2$, $0 \leq r \leq a$.

To reiterate, care must be taken that every term in the small
mass expansion makes sense, either by satisfying the bound
(\ref{Eq_0230}) or by making sure that the offending term is cancelled
by other terms.

\section{\label{Sec_SecIII}
Low-Energy Phase Shifts}

In order to take the small mass limit of $det$ in (\ref{Eq_0226}) we
will need the low-energy phase shifts. From here on it is convenient
to revert to the solutions of

\begin{equation}
\label{Eq_0301}
H_{\pm,l} \psi_{\pm,l} = k^2 \psi_{\pm,l},
\end{equation}

\noindent
where $H_{\pm,l}$ is defined by (\ref{Eq_0204}) and $\psi_{\pm,l}$ are
connected to the regular solutions (\ref{Eq_0211}) of (\ref{Eq_0210}) by

\begin{equation}
\label{Eq_0302}
\psi_{\pm,l}(k,r) = \frac{\varphi_{\pm,l}(k,r)}{\sqrt{r}}.
\end{equation}

\noindent
For any chirality the zero-energy solutions (\ref{Eq_0229}) of
(\ref{Eq_0210}) are related to the zero-energy solutions $\psi^0_l$ of
(\ref{Eq_0301}) by

\begin{equation}
\label{Eq_0303}
\psi^0_l(r) = \frac{R_l(r)}{\sqrt{r}}.
\end{equation}

From (\ref{Eq_0211}), (\ref{Eq_0212}), (\ref{Eq_0223}),
(\ref{Eq_0224}) and (\ref{Eq_0302}), for $r > a$,

\begin{widetext}
\begin{equation}
\label{Eq_0304}
\psi_l(k,r) = 2^{-\half} e^{i\delta_l} e^{i\pi|l|/2}
  (J_W(kr) \, \cos \, \Delta_l - Y_W(kr) \, \sin \, \Delta_l),
\end{equation}
\end{widetext}

\noindent
where $Y_W$ is a Bessel function of the second kind. This holds for
all $l$ and both chiralities except for positive chirality when $l = N$,
which has to be dealt with separately. Then

\begin{equation}
\label{Eq_0305}
\tan \Delta_l =
  \frac{\gamma_l J_W(ka) - ka J'_W(ka)}{\gamma_l Y_W(ka) - ka Y'_W(ka)},
\end{equation}

\noindent
where

\begin{eqnarray}
\label{Eq_0306}
\nonumber
\gamma_l
  &=& (r \, \partial_r \psi_l / \psi_l)_a\\
\nonumber
  &=& (r \, \partial_r \psi^0_l/ \psi^0_l)_a
      - \frac{k^2}{\psi^0_l(a) \psi_l(k,a)}
      \int^a_0 dr \, r \psi^0_l(r) \psi_l(k,r)\\
  &\equiv& \gamma^{(0)}_l + (ka)^2 \gamma^{(2)}_l
      + (ka)^4 \gamma^{(4)}_l + O(ka)^6,
\end{eqnarray}

\noindent
and from (\ref{Eq_0228})

\begin{equation}
\label{Eq_0307}
\psi_l(k,r) = \psi^0_l(r) (1 - (ka)^2 \chi_l(r) + O(ka)^4).
\end{equation}

\noindent
Equations (\ref{Eq_0229}) and (\ref{Eq_0303}) give

\begin{widetext}
\begin{eqnarray}
\nonumber
\gamma^{+ (0)}_l &=&
\begin{cases}
\displaystyle
l - \frac{e\Phi}{2\pi}, & l \geq 0\\[3mm]
\displaystyle
l - \frac{e\Phi}{2\pi}
  + \left(\int^a_0 \frac{dr}{r} \left(\frac{r}{a}\right)^{-2l}
  e^{-2e \varphi(r)} \right)^{-1}, & l < 0
\end{cases}\\[3mm]
\label{Eq_0308}
\gamma^{- (0)}_l &=&
\begin{cases}
\displaystyle
\frac{e\Phi}{2\pi} - l
  + \left(\int^a_0 \frac{dr}{r} \left(\frac{r}{a}\right)^{2l}
  e^{2e \varphi(r)} \right)^{-1}, & l > 0\\[3mm]
\displaystyle
\frac{e\Phi}{2\pi} - l, & l \leq 0,
\end{cases}
\end{eqnarray}
\end{widetext}

\noindent
and (\ref{Eq_0306}), (\ref{Eq_0307}) give, for both chiralities,

\begin{eqnarray}
\label{Eq_0309}
\nonumber
\gamma^{(2)}_l &=& - a^{-2} \int^a_0 dr \, r
  \left( \frac{\psi^0_l(r)}{\psi^0_l(a)} \right)^2\\
\nonumber
\gamma^{(4)}_l &=& a^{-2} \int^a_0 dr \, r
  \left( \frac{\psi^0_l(r)}{\psi^0_l(a)} \right)^2
  (\chi_l(r) - \chi_l(a)).\\
\end{eqnarray}

\noindent
The norms of the square-integrable zero modes are, from (\ref{Eq_0303}), (\ref{Eq_0229})
and (\ref{Eq_0220})

\begin{eqnarray}
\label{Eq_0310}
\nonumber
||\psi^0_l||^2
  &=& \int^\infty_0 dr \, r |\psi^0_l|^2 / a^{2l+2}\\
\nonumber
  &=& \int^a_0 \frac{dr}{a} \left(\frac{r}{a}\right)^{2l+1}
      e^{2e \varphi(r)}\\
  & & + \frac{1}{2(W-1)}, \hspace{5mm} l = 0, ..., N-1.
\end{eqnarray}

\noindent
With (\ref{Eq_0305}), (\ref{Eq_0306}), (\ref{Eq_0308}),
(\ref{Eq_0309}), $e\Phi/2\pi = N + \epsilon >> 1$,
$(ka)^2 << \epsilon$, $(ka)^2 << 1 - \epsilon$ the following
low-energy phase shifts are obtained:

\begin{widetext}
\begin{multline}
\label{Eq_0311}
\Delta^{+}_l
  = - \frac{\pi}{\Gamma^2(W)} \left(\frac{ka}{2}\right)^{2W}
    \left\{
    \frac{2}{||\psi^0_l||^2 (ka)^2} + \frac{1}{W}
    + \frac{1}{(1-W) ||\psi^0_l||^2}\right.\\[3mm]
    \left.
    + \frac{(4 (W-1)^2 (2-W) )^{-1} + 2 \gamma^{(4)}_l}
           {||\psi^0_l||^4}
    + O(ka)^2
    \right\}
    + O(ka)^{4W-4}, l = 0, ..., N-2
\end{multline}

\begin{multline}
\label{Eq_0312}
\Delta^{+}_{N-1}
  = - \frac{\pi}{\Gamma^2 (1 + \epsilon)}
    \left(\frac{ka}{2}\right)^{2 + 2\epsilon}
    \left\{
    \frac{2}{||\psi^0_{N-1}||^2 (ka)^2} + \frac{1}{1 + \epsilon}
    - \frac{1}{\epsilon||\psi^0_{N-1}||^2}\right.\\[3mm]
    \left.
    + \frac{(4 \epsilon^2 (1-\epsilon) )^{-1} + 2 \gamma^{(4)}_{N-1}}
           {||\psi^0_{N-1}||^4}
    \right\}
    - \frac{\pi^2 \cot \pi \epsilon}
           {4 \Gamma^4 (1 + \epsilon)||\psi^0_{N-1}||^4}
    \left(\frac{ka}{2}\right)^{4\epsilon}
    + O(ka)^{6 \epsilon}, 
\end{multline}
\end{widetext}

\noindent
provided $\epsilon > 1 / |\ln (ka)|$;

\begin{widetext}
\begin{equation}
\label{Eq_0313}
\Delta^{+}_N
  = \frac{\pi}{\Gamma^2(1-\epsilon)}
    \left( 2 \int^a_0 \frac{dr}{a} \left(\frac{r}{a}\right)^{2N+1}
    e^{2 e \varphi} + \frac{1}{\epsilon - 1}\right)
    \left(\frac{ka}{2}\right)^{2 - 2\epsilon}
    + O(ka)^{4-4\epsilon},
\end{equation}
\end{widetext}

\noindent
provided $1 - \epsilon > 1 / |\ln(ka)|$, and

\begin{equation}
\label{Eq_0314}
\epsilon > (ka)^2 \int^a_0 \frac{dr}{a}
  \left(\frac{r}{a}\right)^{2N+1} e^{2e \varphi}.
\end{equation}

\noindent
This may seem impossible to satisfy for large $e$, but it turns out
that the integral in (\ref{Eq_0314}) decreases as a power of N (see
Appendix). Continuing,

\begin{widetext}
\begin{alignat}{2}
\label{Eq_0315}
\Delta^{+}_l
  &= \frac{2\pi(ka/2)^{2W+2}}{\Gamma^2 (1+W)}
  \int^a_0 \frac{dr}{a} \left(\frac{r}{a}\right)^{2l+1}
  e^{2e \varphi} +  O(ka)^{2W+4},
  &l =& N+1, N+2, ...,\\ 
\intertext{and}
\label{Eq_0316}
\Delta^{+}_l
  &= \frac{\pi(ka/2)^{2W}}{\Gamma^2(W)}
  \left(
  2 \int^a_0 \frac{dr}{r} \left(\frac{r}{a}\right)^{2|l|}
  e^{-2e \varphi} - \frac{1}{W} \right)
  + O\left[(ka)^{4W}, (ka)^{2W+2}\right],
  \hspace{2mm}
  &l =& -1, -2, ...
\end{alignat}

\noindent
For negative chirality,

\begin{eqnarray}
\label{Eq_0317}
\Delta^{-}_l
  &=&
  - \frac{\pi(ka/2)^{2W}}{\Gamma^2(W+1)}
  \left[2 \int^a_0 \frac{dr}{r} \left(\frac{r}{a}\right)^{2l}
  e^{2e \varphi} + \frac{1}{W}\right]^{-1}
  + O\left[(ka)^{4W}, (ka)^{2W+2}\right], l = 1, ..., N,\\[7mm]
\label{Eq_0318}
\Delta^{-}_l
  &=&
  \frac{\pi(ka/2)^{2W}}{\Gamma^2(W)}
  \left[2 \int^a_0 \frac{dr}{r} \left(\frac{r}{a}\right)^{2l}
  e^{2e \varphi} - \frac{1}{W}\right]
  + O\left[(ka)^{4W}, (ka)^{2W+2}\right], l = N+1, N+2, ...,\\[7mm]
\label{Eq_0319}
\Delta^{-}_l
  &=&
  \frac{2\pi(ka/2)^{2W+2}}{\Gamma^2(W+1)}
  \int^a_0 \frac{dr}{a} \left(\frac{r}{a}\right)^{2|l|+1}
  e^{-2e \varphi} + O(ka)^{2W+4}, l = 0, -1, ...
\end{eqnarray}
\end{widetext}

The negative values of $\Delta^{\pm}_l$ for $l = 1, ... , N$ can be
qualitatively understood as due to the repulsive barrier in $V$
mentioned in Sec. \ref{Sec_SecII}. The apparent poles in
$\Delta^{\pm}_l$ when $W$ is integral disappear when a careful limit
is taken. For example, going back to the basic definition
(\ref{Eq_0305}),

\begin{eqnarray}
\label{Eq_0320}
\nonumber
\lim_{\epsilon \to 0} \Delta^{+}_{N-1}
  &=& \frac{\pi}{2} \left( \ln(ka/2) + \gamma
      - \int^a_0 \frac{dr}{a} \left(\frac{r}{a}\right)^{2N-1}
      e^{2e \varphi} \right)^{-1}\\
\nonumber
  & & + O\left(\frac{1}{\ln^3(ka)}\right),\\
\lim_{W \to 0} \Delta^{-}_l
  &=& \frac{\pi}{2 \ln(ka)} + O\left(\frac{1}{\ln^2(ka)}\right),
\end{eqnarray}

\noindent
where $\gamma$ is Euler's constant. The general rule is that simple
poles in $\Delta^{\pm}_l$ when $W$ is integral are replaced with
logarithms of the type $\ln(ka)$.

The main observation here is the presence of the $(ka)^{-2}$ factors
in $\Delta^{+}_l$ for $l = 0, ..., N-1$ which cause each of the
corresponding partial-wave Green's functions
$\mathcal{G}_{+,l}(me^{i\pi/2},r)$ in (\ref{Eq_0219}) to develop a simple
pole in $m^2$ at the origin. These are of course expected due to the N
square-integrable zero modes of $\mathcal{H}_{+,l}$.

We have learned from this calculation that the precise form of these
phase shifts is necessary if large cancellations are to go through in
the calculation of the determinant. This is further discussed in
Sec. \ref{Sec_SecIV}.

\section{\label{Sec_SecIV}
Small-Mass, Strong-Coupling Expansion of $\text{lndet}$}

Because of the rapid falloff of the low-energy phase shifts
with $l$ the sums and integrals in (\ref{Eq_0226}) can be
interchanged. Using entries 5.54.2 of Ref.~\cite{Gradshteyn65} and
1.12.3.3 of Ref.~\cite{Prudnikov88} one obtains

\begin{widetext}
\begin{alignat}{2}
\label{Eq_0401}
\nonumber
\lefteqn{a^{-2} \,
 \int^{\infty}_a dr \, r \ln \left(\frac{r}{a}\right) K^2_W(mr)}\\[3mm]
\nonumber
  =& \half K_{W+1}(ma) K'_W(ma) - \half K_W(ma) K'_{W+1}(ma)&&\\[3mm]
  &  + \frac{W}{2ma}
  \left[
  K_{W+1}(ma) \pdo{W}K_W(ma) - K_W(ma) \pdo{W}K_{W+1}(ma)
  \right]&&\\[3mm]
\label{Eq_0402}
\nonumber
  =& \frac{\Gamma^2(W-1)}{16} \left(\frac{ma}{2}\right)^{-2W}
  + \frac{\pi}{16 \sin \pi W}
  \left[
  2W \psi(W) + \pi W \cot \pi W
  - 2W \ln \left(\frac{ma}{2}\right) - 2W + 1
  \right]
  \left(\frac{ma}{2}\right)^{-2}&\\[5mm]
  &
  + \frac{\Gamma^2(W)}{8 (1-W) (2-W)^2} \left(\frac{ma}{2}\right)^{2-2W}
  + O[(ma)^{4-2W}, (ma)^0],& l \neq N&\\[3mm]
\label{Eq_0403}
\nonumber
  =& \frac{\Gamma^2(\epsilon - 1)}{16}
  \left(\frac{ma}{2}\right)^{-2\epsilon} + \frac{\pi}{16 \sin \pi\epsilon}
  \left[
  2\epsilon \psi(\epsilon) + \pi \epsilon \cot \pi \epsilon
  - 2 \epsilon \ln \left(\frac{ma}{2}\right) - 2 \epsilon + 1
  \right]
  \left(\frac{ma}{2}\right)^{-2}
  - \frac{\pi}{8 \epsilon \sin \pi \epsilon}&&\\[5mm]
  &
  + \frac{\pi^2}{16\Gamma^2(2 + \epsilon)(\sin \pi \epsilon)^2}
  \left(\frac{ma}{2}\right)^{2\epsilon}
  + \frac{\Gamma^2(\epsilon)}{8(1-\epsilon)(2-\epsilon)^2}
    \left(\frac{ma}{2}\right)^{2-2\epsilon}
  + O[(ma)^{2 + 2\epsilon}, (ma)^{4-2\epsilon}],& l = N&,
\end{alignat}
\end{widetext}

\noindent
where $\psi(z) = \Gamma'(z)/ \Gamma(z)$ and
$0 < \epsilon < 1$. Apparent singularities in (\ref{Eq_0402}) and
(\ref{Eq_0403}) at integral values of $W$ cancel when careful limits
are taken. Also required are the following
expansions~\cite{Abramowitz64}:

\begin{widetext}
\begin{equation}
\label{Eq_0404}
K_W(z) = \half \Gamma(W)(z/2)^{-W}
  \left[
  1 + \frac{(z/2)^2}{1-W} + O(z^4)
  \right]
  - \frac{\pi(z/2)^W}{2\Gamma(W+1) \sin \pi W}
  \left[
  1 + \frac{(z/2)^2}{W+1} + O(z^4)
  \right],
\end{equation}

\noindent
and

\begin{equation}
\label{Eq_0405}
I_W(z) K_W(z) = \frac{1}{2W}
  \left[
  1 + \frac{z^2/2}{1-W^2}
  - \frac{\pi(z/2)^{2W}}{W\Gamma^2(W) \sin \pi W}
  + O(z^4, z^{2W+2})
  \right].
\end{equation}
\end{widetext}

The pole at $m^2 = 0$ in $\mathcal{G}_{+,l} (k=me^{i\pi/2},r)$ from
the factors $(ka)^{-2}$ in $\Delta^{+}_l$ in (\ref{Eq_0311}) and
(\ref{Eq_0312}) make the positive chirality terms for $l = 0,..,N-1$
in (\ref{Eq_0226}) the dominant ones when $ma<<1$. Using
(\ref{Eq_0311})-(\ref{Eq_0319}), (\ref{Eq_0402})-(\ref{Eq_0405}) and
(\ref{Eq_0228}) when it makes sense--as discussed at the end of
Sec. II and below--we obtain from (\ref{Eq_0226})

\begin{widetext}
\begin{multline}
\label{Eq_0406}
\pdo{e} lndet = -2e \int^a_0 dr \, r \varphi(r) B(r)
  + \sum^{N-1}_{l=0} \pdo{e} \ln ||\psi^0_l||^2
  + \frac{\epsilon \Phi}{\pi} \ln(ma)
  - \frac{\Phi}{2\pi}[2\epsilon \psi(\epsilon)
    + \pi \epsilon \cot \pi \epsilon + 1 - 2 \epsilon + 2\epsilon \ln2]\\
  + O[(ma)^{2\epsilon}\ln(ma),(ma)^{2-2\epsilon}\ln(ma),(ma)^2\ln(e\Phi)],
\end{multline}
\end{widetext}

\noindent
provided $|\ln(ma)|^{-1} < \epsilon < 1 - |\ln(ma)|^{-1}$. Recall that
$e\Phi/2\pi = N + \epsilon$.

Regarding the remainder in (\ref{Eq_0406}), there are twelve cases to
consider: positive/negative chirality, regions inside/outside the
range of B, and the angular momentum ranges
$l \leq -1$, $0 \leq l \leq N$, $l \geq N + 1$ for $e\Phi >> 1$. The
terms of order $(ma)^{2\epsilon} \ln(ma)$ and
$(ma)^{2-2\epsilon}\ln(ma)$ come from positive chirality,
$l = N, N-1$ for $r > a$. The term of order $(ma)^2\ln(e\Phi)$ comes from the
$\int^a_0 dr \, \varphi(r) R^2_{\pm,l}(r) \int^a_r ds / R^2_{\pm,l}(s)$
terms in (\ref{Eq_0226}) summed over values of $l$ in the neighborhood
of $-e\Phi/2\pi$. The presence of the factor $\ln(e\Phi)$ is
tentative: there may be subtle cancellations between the positive and
negative chirality sectors that will eliminate the logarithm. All of
the $O(ma)^2$ remainder estimates are based on what we consider the
worst case, namely, $B(r) \geq 0$, which causes $\varphi(r)$ to be
positive and monotonically decreasing for $0 \leq r < a$.

Our second comment on (\ref{Eq_0406}) concerns large individual terms
in the mass expansion when $e\Phi >> 1$. Consider the second term in
(\ref{Eq_0226}) and the ratio $R_{+.l}(m^2,r)/R_{+,l}(m^2,a)$. As
discussed in Sec. \ref{SubSec_SmallMass}, $R_{+,l}$ can exponentially
increase for $e\Phi >> 1$ for $0 \leq l \lesssim e\Phi/2\pi$. However,
this ratio at $m^2 = 0$ $(R_{+}(0,a) = a^{l+\half})$ and its leading
correction $\chi^{+}_l$ in (\ref{Eq_0231}) are cancelled for each $l$
by the third term in (\ref{Eq_0226}). It remains to understand these
cancellations and to verify that they continue at order $(ma)^6$ and
higher orders.

The terms $R^2_{+,l}(r) \int^a_r ds / R^2_{+,l}(s)$ for $0 \le l \le
N$ in (\ref{Eq_0226}) have not been expanded since there is no
apparent cancellation mechanism. We have found that in one of the
worst cases, when $B(r) = B$ for $r < a$ and zero otherwise, these
terms when left unexpanded vanish as $e\Phi \to \infty$. For $l > N$
these terms remain bounded when expanded, and for $l >> e\Phi/2\pi$
their leading $l$ behavior is cancelled by the negative chirality
sector since the distinction between the two chiralities disappears as
$l \to \infty$.

In the exactly solvable case of a magnetic field confined to the
surface of a cylinder the mass-dependent terms remain subdominant when
$e\Phi >> 1$~\cite{Fry95}. The study of the cancellation of large
terms and the vanishing of ratios of large terms when $e\Phi \to
\infty$ is still at a preliminary stage. The control of these terms
has much to teach us about the nonperturbative structure of $lndet$.

Finally, we have previously shown that for $e\Phi$ fixed and $ma << 1$

\begin{equation}
\label{Eq_0407}
lndet = \frac{|e\Phi|}{2\pi} \ln(ma) + R(m),
\end{equation}

\noindent
where $\lim_{m=0}(R/ln(ma)) = 0$~\cite{Fry00b}. Now consider the case when
$\epsilon = 0$ and $e\Phi/2\pi = N$. Then the dominant mass-dependent
term in (\ref{Eq_0226}) for $ma << 1$ occurs at $l = N-1$, $r > a$:

\begin{widetext}
\begin{equation}
\label{Eq_0408}
\pdo{e}lndet_{N-1}
  = \frac{2m^2\Phi}{\pi^2}
    (\Delta^{+}_{N-1} + i\Delta^{+ 2}_{N-1} - \Delta^{-}_{N-1} + ...)
    \int^{\infty}_a dr \, r \ln\left(\frac{r}{a}\right) K^2_1(mr), 
\end{equation}
\end{widetext}

\noindent
where $\Delta^{\pm}_{N-1}$ are continued to $k = me^{i\pi/2}$. From
(\ref{Eq_0317}), (\ref{Eq_0320}) and

\begin{widetext}
\begin{equation}
\label{Eq_0409}
a^{-2} \int^{\infty}_a dr \, r \ln\left(\frac{r}{a}\right) K^2_1(mr)\\[3mm]
  = \frac{[\ln(ma/2) + \gamma]^2 + \ln (ma/2) + \gamma + 1}{2(ma)^2}
  + \frac{1}{4} \ln\left(\frac{ma}{2}\right)
  + \frac{\gamma}{4} - \frac{3}{8} + O[(ma)^2\ln^2(ma)],
\end{equation}
\end{widetext}

\noindent
one gets

\begin{equation}
\label{Eq_0410}
\pdo{e} lndet_{N-1} = \frac{\Phi}{2\pi} \ln(ma) + O(1),
\end{equation}

\noindent
in accord with (\ref{Eq_0407}).

Next, consider the case when $e\Phi/2\pi = N + \epsilon$,
$0 < \epsilon \leq 1$. As $\epsilon \to 1$ a pole at $m^2 = 0$ begins
to develop in $\mathcal{G}_{+,N}$ and
$\Delta^{+}_N(k) \sim \pi / [2\ln(ka)]$. For $e\Phi/2\pi = N+1$ we
find

\begin{equation}
\label{Eq_0411}
\pdo{e} lndet_N = \frac{\Phi}{2\pi}\ln(ma) + O(1),
\end{equation}

\noindent
again in accord with (\ref{Eq_0407}). Moreover, the same result is
obtained in the limit $\epsilon \to 1$.

In the interval $0 < \epsilon < 1$ the $(\epsilon\Phi/\pi)\ln(ma)$
term in (\ref{Eq_0406}) comes from the $l=N$, $r > a$ contribution to
$\partial lndet / \partial e$. This term contradicts (\ref{Eq_0407})
which was derived by holding $e\Phi$ fixed and letting $ma \to
0$. Here we are setting $ma << 1$, then letting $e\Phi$ increase
indefinitely. By taking limits in this way the $\ln(ma)$ term becomes
an infinitesimal addition to $lndet$ when compared to its growth due
to the pileup of normalizable zero modes as $e\Phi$ increases, as we
will see in Sec. \ref{Sec_SecV}. For the present it is assumed that
there are other infinitesimal terms not yet found that will result in
the shift $(\epsilon \Phi / \pi)\ln(ma) \to (\Phi/2\pi)\ln(ma)$ in the
range of $\epsilon$ indicated.

We are confident that (\ref{Eq_0407}) is the leading mass-dependent term
in $lndet$, and it will accordingly be added on to our strong coupling
result for $lndet$ in Sec. \ref{Sec_SecV}.

\section{\label{Sec_SecV}
Small-Mass, Strong-Coupling Limit of $\text{lndet}$}

Up to now we have assumed that $B(r)$ is square-integrable,
centrally symmetric and finite-ranged. Further analytic analysis of
(\ref{Eq_0406}) requires additional assumptions, namely $B(r) \ge 0$
with continuous first and second derivatives. Then we can show that
for $e\Phi \to \infty$, the first term in (\ref{Eq_0406}) is cancelled
by the zero modes contributing to the second term.

The demonstration is straightforward. Refer to (\ref{Eq_0406}),
(\ref{Eq_0303}), the first lines of (\ref{Eq_0310}) and
(\ref{Eq_0229}) and obtain

\begin{widetext}
\begin{equation}
\label{Eq_0501}
\sum^{N-1}_{l=0} \pdo{e}\ln||\psi^0_l||^2
  = 2 \int^{\infty}_0 dr \, r \varphi(r)
    \sum^{N-1}_{l=0}
    \frac{r^{2l} e^{2e\varphi(r)}}
         {\int^{\infty}_0 ds \, s^{2l+1} e^{2e\varphi(s)}}.
\end{equation}
\end{widetext}

\noindent
Now make use of the following theorem of Erd\"{o}s~\cite{Erdos93},
specialized here to the case of central symmetry: Let $B(r)\ge0$ be a
compactly supported magnetic field with a continuous first
derivative. Define the ground-state density function

\begin{equation}
\label{Eq_0502}
P(r) = \sum^{N-1}_{l=0}
  \frac{r^{2l} e^{2e\varphi(r)}}
       {\int^{\infty}_0 ds \, s^{2l+1} e^{2e\varphi(s)}}.
\end{equation}

\noindent
Then $P(r)/e$ converges to $B(r)$ in $L^p$  for any $1 \le p < \infty$
as $e \to \infty$. According to this theorem

\begin{equation}
\label{Eq_0503}
\sum^{N-1}_{l=0} \pdo{e}\ln ||\psi^0_l||^2
  = 2e \int^a_0 dr \, r \varphi(r) B(r) + R(e),
\end{equation}

\noindent
for $e\Phi >> 1$ and where $\lim_{e \to \infty} R(e)/e = 0$. The
$r$-integral in (\ref{Eq_0501}) cuts off due to the finite range of
$B$. Hence, (\ref{Eq_0503}) leads to the promised cancellation in
(\ref{Eq_0406}).

The really interesting question now is what is the remainder in
(\ref{Eq_0503})? Erd\"{o}s' theorem is not yet sharp enough to state
what it is. It had better be negative to be in accord with the
diamagnetic upper bound in (\ref{Eq_0230}). In the Appendix we
investigate this problem by the method of steepest descents assuming
$B(r)>0$ with two alternative sets of boundary conditions: $B(a)=0$,
$\lim_{r \to a-} B'(r) < 0$, and $B(a)>0$. The result in both cases is

\begin{widetext}
\begin{equation}
\label{Eq_0504}
\lim_{|e\Phi| >> 1} \lim_{ma << 1} lndet
  = -\frac{|e\Phi|}{4\pi} \ln\left(\frac{|e\Phi|}{(ma)^2}\right)
    + O(|e\Phi|,(ma)^2|e\Phi|\ln(|e\Phi|)).
\end{equation}
\end{widetext}

\noindent
The case when $eB < 0$ is the mirror image of the $eB > 0$ case, and so
we need only insert absolute value signs to cover both cases. As
discussed in Sec. \ref{Sec_SecIV}, we have inserted the mass-dependent
term from Ref.~\cite{Fry00b}. Comparing (\ref{Eq_0504}) with the constant
field result

\begin{equation}
\label{Eq_0505}
lndet = - \frac{eBV}{4\pi} \ln\left(\frac{eB}{m^2}\right) + O(eB),
\end{equation}

\noindent
we see that they are formally in accord on setting
$V = \pi a^2 \to \infty$. Of course we cannot say anything about the
remaining mass-dependent terms in (\ref{Eq_0504}) in this limit.

The minus sign in (\ref{Eq_0504}) is a reflection of the paramagnetism
of charged fermions in a magnetic field. This is most clearly seen
with Schwinger's proper time definition of the
determinant~\cite{Schwinger51}, namely

\begin{widetext}
\begin{equation}
\label{Eq_0506}
lndet = \half \int^{\infty}_0 \frac{dt}{t} e^{-tm^2} Tr
  \left[
  e^{-P^2 t}
  \right.
  \left.
  - \exp\left\{-[(P-eA)^2 - \sigma_3 B]t\right\} \;
  \right].
\end{equation}
\end{widetext}

\noindent
Noting the minus sign in (\ref{Eq_0504}), (\ref{Eq_0506})
indicates that on average the spectrum of the Pauli operator is
lowered by $B$ relative to the field-free case. Therefore, the current
usage of "diamagnetic" bound to describe the right-hand side of
(\ref{Eq_0230}) is a misnomer. The factor $|e\Phi|$ in (\ref{Eq_0504})
multiplying the logarithm is related to the counting of zero
modes. More will be said about the physics of (\ref{Eq_0504}) in
Sec. \ref{Sec_SecVI}.

The discussion of the remainder in (\ref{Eq_0406}) in
Sec. \ref{Sec_SecIV} means that we cannot rule out the subdominant
term $(ma)^2 |e\Phi| \ln(|e\Phi|)$ in (\ref{Eq_0504}); more detailed
analysis is required to exclude the $\ln(|e\Phi|)$ factor.

The remarkable thing about (\ref{Eq_0504}) is that the limit is
universal for a broad class of fields. Since it only depends on a
global property of the background magnetic field--its total flux--we
suspect that (\ref{Eq_0504}) is also the limit in the general case of
non-central, square-integrable fields.

Finally, the case of zero-flux background fields has not been
considered in the literature to the author's knowledge except for the
case of massless QED$_2$ on a torus~\cite{Sachs92} and a
sphere~\cite{Jayewardena88}. Our limit seems to indicate that when
$\Phi = 0$ there are no square-integrable zero modes and hence no
mechanism to cancel the first term in (\ref{Eq_0406}).  In this case
one might suppose that it is this term--the Schwinger term--that is
dominant in the small-mass, strong-coupling limit. This is the result
in~\cite{Sachs92}.


\section{\label{Sec_SecVI}
Duality}

The purpose of this section is to relate the Euclidean determinant of
QED$_2$ and some of the results of the previous sections to physics in
four dimensions. The term duality as used in this section is distinct
from Olive-Montonen electric-magnetic duality~\cite{Montonen77}. It is
rather a duality closely related to the analyticity of the one-loop
effective action of QED in two and four dimensions.

The Euclidean determinants in QED$_4$ and QED$_2$ for the background
magnetic field $B = (0,0,B(x_1, x_2))$ are related by

\begin{widetext}
\begin{equation}
\label{Eq_0601}
-2\pi \pdo{m^2} lndet_{\text{QED}_4}
  = L_3 L_4 lndet_{\text{QED}_2}
    + \frac{L_3 L_4 ||B||^2 e^2}{12\pi m^2},
\end{equation}
\end{widetext}

\noindent
where $||B||^2 = \int dx_1 dx_2 B^2(x_1, x_2)$, $L_3 L_4$ is the
volume of the space-time box for $x_3$ and $x_4$, and on-shell charge
renormalization is used~\cite{Fry92}. Hence $B$ must be at least
square-integrable in what follows.  Assuming one can rotate energy
contours in the usual way, continue $lndet_{\text{QED}_4}$ to the
Lorentz metric by letting $\gamma_4 \to i \gamma_0$,
$x_4 \to e^{i(\pi/2 - \epsilon)}t$, $\epsilon \to 0+$ and $L_4 \to iT$.
On the right det$_{\text{QED}_2}$ remains a Euclidean determinant and
so (\ref{Eq_0601}) now becomes

\begin{widetext}
\begin{equation}
\label{Eq_0602}
-2 \pi \pdo{m^2} lndet^L_{\text{QED}_4}(B)
  = i L_3 \, lndet^E_{\text{QED}_2}(B)
    + \frac{i L_3 T ||B||^2 e^2}{12 \pi m^2},
\end{equation}
\end{widetext}

\noindent
with the superscripts $E$ and $L$ denoting Euclidean and Lorentz metrics,
respectively. Therefore, given $det^E_{\text{QED}_2}(B)$ we can
calculate $det^L_{\text{QED}_4}(B)$ for a general unidirectional
magnetic field $B(\mathbf{r})$ by integrating (\ref{Eq_0602}) over
$m^2$  as described in Ref.~\cite{Fry92}.

Now make the duality transformation from the static magnetic field
$B(x_1, x_2)$ to the functionally equivalent electric field $E(x_3,t)$
by letting

\begin{equation}
\label{Eq_0603}
\mathbf{A} = (A_1(x_1,x_2), A_2(x_1,x_2), 0) \to (0,0,A_3(x_3,t)),
\end{equation}

\noindent
with $\mathbf{\nabla x A} = B(x_1, x_2) \mathbf{\hat{k}}$,
$\mathbf{E} = -\dot{A}_3 \mathbf{\hat{k}} = B(x_3,t) \mathbf{\hat{k}}$ and

\begin{equation}
\label{Eq_0604}
A_3(x_3,t) = - \int^t_{t_0} ds \, B(x_3, s).
\end{equation}

\noindent
A change in $t_0$ in (\ref{Eq_0604}) results in a gauge transformation
and does not affect the determinant. This duality transformation is
implemented by the replacement $B(x_1,x_2) \to e^{-i\pi/2} E(x_3,x_4)$
in det$^E_{\text{QED}_4}$, det$^E_{\text{QED}_2}$ and $||B||$ in
(\ref{Eq_0601}) and the coordinate/momentum relabeling
$1 \leftrightarrow 3$, $2 \leftrightarrow 4$, followed by continuation
to the Lorentz metric, including $b \to e^{i\pi/2} \tau$, where $b$ is
the range of $B$ in the $x_2$-direction, and $2\tau$ is the duration
of the electric pulse $E(x_3,t)$. An example is given in
(\ref{Eq_0607}) below. If $B$ has more than one range parameter in the
$x_2$-direction then all of them must be continued as $b$. The rule
$B \to e^{-i\pi/2}E$ in going from the Euclidean metric back to the
Lorentz metric is a consequence of the definition of $E$ above and the
rotation $x_4 \to e^{i(\pi/2 - \epsilon)}t$. Ultimately it is rooted
in the fundamental prescription $m^2 \to m^2 - i\epsilon$. Then
(\ref{Eq_0601}) becomes

\begin{widetext}
\begin{equation}
\label{Eq_0605}
-2\pi \pdo{m^2} lndet^{E \to L}_{\text{QED}_4} (B \to e^{-i\pi/2}E)
  = L_1 L_2 lndet^{E \to L}_{\text{QED}_2} (B \to e^{-i\pi/2} E)
    - \frac{i L_1 L_2 ||E||^2 e^2}{12 \pi m^2}.
\end{equation}
\end{widetext}

\noindent
As an example consider the last terms in (\ref{Eq_0601}) and
(\ref{Eq_0605}) for the case of a magnetic field in a closed region
with two range parameters:

\begin{eqnarray}
\label{Eq_0606}
\nonumber
B(x_1, x_2) &=& B f\left(\frac{x_1}{a}, \frac{x_2}{b}\right),\\
E(x_3,t)    &=& B f\left(\frac{x_3}{a}, \frac{t}{\tau}\right),
\end{eqnarray}

\noindent
with $B(x_1 = \pm a, x_2 ) = B(x_1, x_2 = \pm b) = 0$. Following the
above rules

\begin{eqnarray}
\label{Eq_0607}
\nonumber
||B||^2
  &=& B^2 \int^a_{-a} dx_1 \int^{b g_2(x_1/a)}_{b g_1(x_1/a)}
      dx_2 f^2 \left(\frac{x_1}{a}, \frac{x_2}{b}\right)\\
\nonumber
  &\to& -B^2 \int^a_{-a} dx_3 \int^{i \tau g_2(x_3/a)}_{i\tau g_1(x_3/a)}
      dx_4 f^2 \left(\frac{x_3}{a}, \frac{x_4}{e^{i\pi/2} \tau}\right)\\
\nonumber
  &=& -i B^2 \int^{a}_{-a} dx_3 \int^{\tau g_2(x_3/a)}_{\tau g_1(x_3/a)}
      dt f^2 \left( \frac{x_3}{a}, \frac{t}{\tau}\right)\\
  &=& -i ||E||^2,
\end{eqnarray}

\noindent
where $x_2 = b g_i(x_1/a)$, $t = \tau g_i(x_3/a)$, $i = 1, 2$ define
the boundaries of $B$ and $E$.

Equation (\ref{Eq_0605}) may seem to give nothing new, at least when
developed in a power-series expansion in $E$. Its real power enters when
$lndet^E_{\text{QED}_2}(B)$ is known nonperturbatively as we will now
see. Defining the one-loop Lorentz metric effective action by
$S_\text{eff} = -i \, lndet$, (\ref{Eq_0605}) gives

\begin{widetext}
\begin{equation}
\label{Eq_0608}
-2\pi \pdo{m^2} S^{\text{QED}_4}_{\text{eff}}(E)
  = - i L_1 L_2 \, lndet^{E \to L}_{\text{QED}_2} (B \to e^{-i\pi/2} E)\\
    - \frac{L_1 L_2 ||E||^2 e^2}{12 \pi m^2}.
\end{equation}
\end{widetext}

\noindent
As an example, consider the finite-range magnetic field

\begin{equation}
\label{Eq_0609}
B(x_1, x_2) = \left( 1 - \frac{x^2_1 + x^2_2}{a^2} \right) B,
  x^2_1 + x^2_2 \le a^2,
\end{equation}

\noindent
and the corresponding electric pulse

\begin{equation}
\label{Eq_0610}
E(x_3, t) = \left( 1 - \frac{x^2_3 + (ct)^2}{a^2} \right) E,
  x^2_3 + (ct)^2 \le a^2,
\end{equation}

\noindent
where $B$ and $E$ are constants, $\Phi = \pi a^2 B/2$ and
$c \tau = a$. Both fields are directed along the $z$-axis. For
$ma \to 0$ and $e \Phi >> 1$ we found the result (\ref{Eq_0504}) for
$lndet^E_{\text{QED}_2}(B)$. Then following the above rules set

\begin{eqnarray}
\nonumber
\lefteqn{lndet^{E\to L}_{\text{QED}_2} (B \to e^{-i\pi/2} E)}\\
\nonumber
  &=& -\frac{(e\pi a/2)(e^{i\pi/2}\tau)(e^{-i\pi/2}E)}{4\pi}
      \ln \left(\frac{e\pi e^{-i\pi/2}E}{m^2}\right)\\
\nonumber
  & & \hspace{10mm} + O(eE)\\[2mm]
\label{Eq_0611}
  &=& -\frac{e\pi a \tau E}{8\pi} \ln\left(\frac{eE}{m^2}\right)
      + O(eE),
\end{eqnarray}

\noindent
where corrections of $O((ma)^2)$ have been ignored. Substituting
(\ref{Eq_0611}) in (\ref{Eq_0608}) gives for $eE >> m^2$

\begin{widetext}
\begin{equation}
\label{Eq_0612}
2\pi \pdo{m^2} \text{Im} \, S^{\text{QED}_4}_{\text{eff}}
  = -\frac{e\pi a \tau L_1 L_2 E}{8\pi} \ln\left(\frac{eE}{m^2}\right)
    + O(eE).
\end{equation}
\end{widetext}

\noindent
As far as we know there is nothing in the literature to direcly check
(\ref{Eq_0612}) with, or any other class of electric two-variable
pulses.

The minus sign in (\ref{Eq_0612}) is universal for the class of fields
and their dual pulses considered in this paper. We now see that the
physically reasonable result that the pair production probability
$1 - \text{exp}(-2 \, Im \, S_{\text{eff}})$ decreases with increasing
fermion mass depends on the paramagnetism of charged fermions in a
magnetic field, as indicated by the minus sign in (\ref{Eq_0504}) and
discussed afterwards. We take this as direct physical evidence for the
validity, at least in the strong-coupling, low-mass domain, of the
"diamagnetic" bound on the Euclidean determinant, namely
det$^E_{\text{QED}_2} \le 1$.

The diamagnetic bound also holds in the perturbative domain of large
mass and weak coupling since the power series expansion of
$lndet^{E}_{\text{QED}_2}$ is asymptotic and the overall sign of the
second-order term is negative~\cite{Fry00a}.

A mechanical device that would simulate the pulses implied by the
duality transforms on centrally symmetric magnetic fields would be
two parallel conducting plates of large extent initially very close
together, then pulled apart and then pushed together again. These
plates have the unusual property of having opposite surface-charge
densities varying with time and their spatial separation.

Duality has been considered recently by Dunne and Hall~\cite{Dunne98}
for nonconstant fields in their study of the exactly solvable
single-variable magnetic field $B(x)=B \, \text{sech}^2(x/\lambda)$.
Although the asymptotic boundary conditions are different in the
magnetic and electric field cases, they allow the analytic
continuations required for duality in this example. In a later
paper~\cite{Dunne99} they go beyond exactly solvable background fields
by using a WKB approach to approximate the spectrum of the Pauli
operator $(\slashletter{P} - e\slashletter{A})^2$. The authors are
aware that such an approach cannot prove duality in the
single-variable case, but it does give an insight into just how
nontrivial duality is. Presumably the final justification of duality
in both the one- and two-variable cases is the validity of the Wick
rotation in the presence of external fields.

The question arises as to whether there is a duality transformation of
the type $B(x_1,x_2) \to e^{-i\pi/2} E(x_1, x_2)$, where $E$ is
directed along the third axis. The answer is no except for the special
case when $E(x_1 ,x_2)$ is constant within the boundary parallel to
the direction of the field. Otherwise, the Bianchi identity excludes
such fields. So for $B$ constant over a finite spatial region, duality
takes the simple form, from (\ref{Eq_0602}),

\begin{widetext}
\begin{equation}
\label{Eq_0613}
-2\pi \pdo{m^2} S^{\text{QED}_4}_{\text{eff}}
  (B \to e^{-i\pi/2} E)
  = L_3 T \, lndet^E_{\text{QED}_2} (B \to e^{-i\pi/2} E)
  - \frac{L_3 T ||E||^2 e^2}{12 \pi m^2}. 
\end{equation}
\end{widetext}

\noindent
The determinant det$_{\text{QED}_2}$ retains its Euclidean metric
since the background field is static. For the case of a circular
boundary of radius $a$ (\ref{Eq_0613}) can be checked since there is a
reliable semiclassical approximation that is valid for $a^2eE >> \pi$,
namely~\cite{Martin88}

\begin{widetext}
\begin{equation}
\label{Eq_0614}
Im \, S^{\text{QED}_4}_{\text{eff}}
  = \frac{L_3 T e^2 E^2}{8 \pi^3}
  \sum^{\infty}_{n=1} \frac{1}{n^2} e^{\textstyle{\frac{-n\pi m^2}{eE}}}
  \left[
  \pi a^2 - \pi^2 a \left(\frac{n}{eE}\right)^{\half}
  \text{erf} \, \left(a\left(\frac{eE}{n\pi}\right)^{\half}\right)
  + \frac{n\pi^2}{eE} (1 - e^{-a^2 e E / n\pi})
  \right].
\end{equation}
\end{widetext}

\noindent
Then for $a^2 e E >> \pi >> m^2 a^2$,

\begin{equation}
\label{Eq_0615}
-2\pi \pdo{m^2} Im \, S^{\text{QED}_4}_{\text{eff}}
  = \frac{L_3 T \pi a^2 e E}{4 \pi}[\ln(eE/m^2) + O(1)],
\end{equation}

\gap
\noindent
which agrees by inspection with (\ref{Eq_0613}) when combined with
(\ref{Eq_0504}), taking $e\Phi > 0$ and letting
$\Phi = \pi a^2 B \to \pi a^2 e^{-i\pi/2} E$.

\section{\label{Sec_SecVII}
Summary}

An exact representation of the Euclidean fermion determinant in two
dimensions for centrally symmetric, finite-ranged Abelian background
gauge fields has been obtained that depends only on the interior
partial-wave functions and scattering phase shifts continued to the
upper $k$-plane by setting $k = me^{i\pi/2}$, where $m$ is the fermion
mass. In the nonperturbative limit of small fermion mass these are
known explicitly, thereby making the determinant amenable to numerical
analysis. For the sequence of limits of small fermion mass followed by
strong coupling we have been able to obtain the explicit asymptotic
limit of the determinant when the background field is unidirectional
and nonvanishing except on its boundary. The result is universal,
depending only on the two-dimensional chiral anomaly $e\Phi/2\pi$. It
should be an easy task to obtain the determinant's asymptotic limit
for fluctuating magnetic fields since one only needs to numerically
evaluate the second term in (\ref{Eq_0406}). These results should be a
useful nonperturbative check on lattice algorithms for fermion
determinants when the output is extrapolated to infinite volume and
zero lattice spacing.

By extending the concept of duality to two variables we have been able
to relate the Euclidean determinant in two dimensions for a wide class
of background magnetic fields to the pair production probability in
four dimensions for a related class of electric pulses. We have also
connected the "diamagnetic" bound on the Euclidean two-dimensional
determinant to the negative sign of
$\partial ImS_{\text{eff}} / \partial m^2$ in four dimensions, thereby
providing a physical basis for this bound in the strong-coupling,
small-mass limit.

Central to this work was the ability to count zero modes in two
dimensions. Further analytic progress in three and four dimensions
will be hindered, if not blocked, until there are theorems for
counting zero modes. In four dimensions more is needed than just the
difference of positive and negative chirality zero modes, while in
three dimensions there may be some as yet undiscovered topological
invariant that will count them.

\begin{acknowledgments}
The author would like to thank G. Dunne, L. Erd\"{o}s and C. Lang for
helpful correspondence, and M. Peardon, S. Ryan and I. Sachs for
helpful discussions.
\end{acknowledgments}

\appendix*

\section{}

\setcounter{equation}{0}
\renewcommand{\theequation}{A\arabic{equation}}

Here we will derive the asymptotic limit (\ref{Eq_0504}). Referring to
(\ref{Eq_0310}) let

\begin{equation}
\label{Eq_A01}
I = \int^a_0 \frac{dr}{a}
  \left(\frac{r}{a}\right)^{2l+1} e^{2e \varphi(r)}.
\end{equation}

\noindent
Then,

\begin{eqnarray}
\nonumber
\lefteqn{\pdo{e} \ln ||\psi^0_l||^2}\\
\nonumber
  &=& \pdo{e} \ln I - \frac{\Phi}{2\pi (W-1)}
  + \frac{\Phi}{2\pi} \left(W - 1 + \frac{1}{2I}\right)^{-1}\\
\label{Eq_A02}
  & & - \frac{1}{2I} \left(W-1 + \frac{1}{2I}\right)^{-1} \pdo{e} \ln I.
\end{eqnarray}

\noindent
Consider the first term in (\ref{Eq_A02}). Referring to (\ref{Eq_0406})
consider

\begin{equation}
\label{Eq_A03}
\sum^{N-1}_{l=0} \pdo{e} \ln I
  = \sum^{\Lambda}_{l=0} \pdo{e} \ln I
    + \sum^{N-1}_{l=\Lambda + 1} \pdo{e} \ln I,
\end{equation}

\noindent
where $\Lambda >> 1$ and where for $l \leq N-1$,
$W = e\Phi / (2\pi) - l = N + \epsilon - l$. Refer to the first sum
in (\ref{Eq_A03}). By inspection of (\ref{Eq_A01}),
$I(l=0) = O(e^{2eM})$, where $M = \max \varphi(r)$, $0 \le r \le a$,
with $\varphi(r)$ given by (\ref{Eq_0220}). Hence,
$\partial \ln I / \partial e = O(M)$. For $l = O(N - \gamma N)$, where
$\gamma << 1$ we find later on in (\ref{Eq_A40}) with
$m = N-l-1 = O(\gamma N)$ that
$\partial\ln I / \partial e = O(\sqrt{\gamma})$. These two results
indicate that $I$ has exponential growth in $e$ for this range of
$l$. Thus, $\partial\ln I / \partial e = O(1)$ or less for
$0 \le l \le \Lambda$ and

\begin{equation}
\label{Eq_A04}
\sum^{\Lambda}_{l=0} \pdo{e} \ln I = O(\Lambda).
\end{equation} 

Now for the second sum in (\ref{Eq_A03}). For $\Lambda$ large enough
we can use the method of steepest descents to calculate $I$ except
near the point $l=N-1$. Referring to (\ref{Eq_A01}), let

\begin{equation}
\label{Eq_A05}
f(r) = (2l+1) \ln \left(\frac{r}{a}\right) + 2e \varphi(r).
\end{equation}

\noindent
Assume $B(r) > 0$ so that $\Phi(r)$ given by (\ref{Eq_0205}) is
monotonically increasing with $r$. Then $f(r)$ is maximized at point
$r^{*}$ for which

\begin{equation}
\label{Eq_A06}
l + \half = e\Phi(r^{*}) / 2\pi,
\end{equation}

\noindent
since $f''(r^{*}) = -2eB(r^{*}) < 0$. Hence for $l>>1$,

\begin{equation}
\label{Eq_A07}
I = \sqrt{\frac{2\pi}{a^2|f''(r^{*})|}} e^{f(r^{*})} (1 + O(1/N)).
\end{equation}

\noindent
To calculate the point $r^{*}$ for each admissible $l$, note that for
$l \to N$, $r^{*} \to a$. So expand the right-hand side of
(\ref{Eq_A06}) about $r^{*} = a$ by setting $r^{*} = a (1 - \delta)$. 
Let $l = N-m-1$, $m >> 1$, $m << N$. Assuming $B(r)$ has continuous
first and second derivatives with $B(a) = 0$ and $B'(a) < 0$ then
$\delta = (2 / (a^3 |B'(a)|))^{\half} (m/e)^{\half} + O(m/e)$ and

\begin{equation}
\label{Eq_A08}
f(r^{*})
  = \frac{4}{3} \left(\frac{\Phi}{\pi a^3 |B'(a)|}\right)^{\half}
    \frac{m^{\frac{3}{2}}}{\sqrt{N}} + O\left(\frac{m}{N}\right)^{\half},
\end{equation}

\noindent
and

\begin{equation}
\label{Eq_A09}
|f''(r^{*})|
  = 4 \left(\frac{\pi|B'(a)|}{a\Phi}\right)^{\half} \sqrt{mN}
    \left(1 + O\left(\frac{m}{N}\right)^{\half}\right).
\end{equation}

\noindent
Inserting (\ref{Eq_A08}), (\ref{Eq_A09}) in (\ref{Eq_A07}) gives for
$m << N$

\begin{widetext}
\begin{equation}
\label{Eq_A10}
I = \left(\frac{\pi \Phi}{4 a^3 |B'(a)|}\right)^{\frac{1}{4}}
    (mN)^{-\frac{1}{4}}
  \exp
  \left[
  \frac{4}{3} \left(\frac{\Phi}{\pi a^3|B'(a)|}\right)^{\half}
  \frac{m^{\frac{3}{2}}}{\sqrt{N}} + O\left(\frac{m}{N}\right)^{\half}
  \right]
  \left[1 + O\left(\frac{m}{N}\right)^{\half}\right]. 
\end{equation}
\end{widetext}

\noindent
By definition (\ref{Eq_A01}), $\partial I / \partial m > 0$ for
$0 \le m \le N-1$. This will be true for the estimate (\ref{Eq_A10})
provided

\begin{equation}
\label{Eq_A11}
m > \left(\frac{\pi a^3 |B'(a)|}{64 \Phi} \right)^{\frac{1}{3}}
  N^{\frac{1}{3}} \equiv CN^{\frac{1}{3}},
\end{equation}

\noindent
in addition to $m << N$.

Now return to the second sum in (\ref{Eq_A03}) and write it as the
following sum using (\ref{Eq_A07}):

\begin{eqnarray}
\nonumber
\lefteqn{\sum^{N-1}_{l=\Lambda+1} \pdo{e} \ln I}\\
\label{Eq_A12}
  &=& \left(\sum^{N-CN^{\frac{1}{3}}}_{l = \Lambda + 1}
      + \sum^{CN^{\frac{1}{3}}}_{m=0}\right) \pdo{e} \ln I\\
\nonumber
  &=& \sum^{N-CN^{\frac{1}{3}}}_{l=\Lambda + 1} \pdo{e} f(r^{*}_l)
      - \half \sum^{N-CN^{\frac{1}{3}}}_{l=\Lambda+1}
      \pdo{e} |f''(r^{*}_l)|\\
\label{Eq_A13}
  & & + \sum^{CN^{\frac{1}{3}}}_{m=0} \pdo{e} \ln I
      + O\left(\frac{1}{N^2}\right).
\end{eqnarray}

\noindent
Consider the first term in (\ref{Eq_A13}). We need not rely on
(\ref{Eq_A08}) yet because (\ref{Eq_A07}) holds irrespective of where
the roots $r^{*}_l$ of (\ref{Eq_A06}) lie in $(0,a)$. The important
point is that they are closely spaced over the entire interval $(0,a)$
for $e\Phi / 2\pi >> 1$ and for $\Phi(r)$ monotonically increasing
with $r$. Hence, the $r^{*}_l$ can be considered to be nearly
continuous across $(0,a)$ for $l$ in the range indicated with

\begin{equation}
\label{Eq_A14}
dl = \frac{e}{2\pi} \odo{r^{*}_l} \Phi(r^{*}_l) \, dr^{*}_l
   = eB(r^{*}_l) r^{*}_l dr^{*}_l.
\end{equation}

\noindent
Referring to (\ref{Eq_A05}), (\ref{Eq_A06}) and (\ref{Eq_0220}),

\begin{equation}
\label{Eq_A15}
\pdo{e}f(r^{*}_l) = 2 \varphi(r^{*}_l),
\end{equation}

\noindent
and so

\begin{eqnarray}
\nonumber
\lefteqn{\sum^{N - CN^{\frac{1}{3}}}_{l=\Lambda + 1} \pdo{e} f(r^{*}_l)}\\
\nonumber
  &=& 2 \sum^{N - CN^{\frac{1}{3}}}_{l=\Lambda + 1} \varphi(r^{*}_l)\\
\label{Eq_A16}
  &=& 2e \int^a_0 dr^{*} \, r^{*} B(r^{*}) \varphi(r^{*}) + O(1).
\end{eqnarray}

\noindent
When (\ref{Eq_A16}), (\ref{Eq_A13}), (\ref{Eq_A02}) are combined we
already see the promised cancellation of the first term in
(\ref{Eq_0406}), as guaranteed by Erd\"{o}s' theorem~\cite{Erdos93}. We
now turn to the calculation of the remainder.

Consider the second sum in (\ref{Eq_A13}) and break it up into two
sums:

\begin{widetext}
\begin{equation}
\label{Eq_A17}
\sum^{N-CN^{\frac{1}{3}}}_{l=\Lambda+1}
  \pdo{e} \ln|f''(r^{*}_l)|
  = \left(
  \sum^{(1 - \gamma)N}_{l=\Lambda+1}
  + \sum^{N-CN^{\frac{1}{3}}}_{l=(1-\gamma)N}
  \right)
  \pdo{e} \ln |f''(r^{*}_l)|,
\end{equation}
\end{widetext}

\noindent
where $\gamma << 1$. Now deal with the first sum and recall
$f''(r^{*}_l) = -2eB(r^{*}_l)$. From (\ref{Eq_A06}) for
$l = \Lambda + 1$,
$\Phi(r^{*}_l)/\Phi = (\Lambda + \half)/(N + \epsilon)$ which implies
$r^{*}_l \gtrsim 0$ for $N >> \Lambda$, and hence
$f''(r^{*}_l) \simeq -2eB(0)$. For the upper limit
$l = (1 - \gamma)N$, (\ref{Eq_A06}) gives
$\Phi(r^{*}_l)/\Phi = 1 - \gamma + O(1/N)$ and hence
$r^{*}_l \lesssim a$. So

\begin{eqnarray}
\nonumber
f''(r^{*}_l)
  &=& -2eB(a) - 2eB'(a) (r^{*}_l - a) + O(r^{*}_l - a)^2\\
\label{Eq_A18}
  &=& -2 e|B'(a)|(a - r^{*}_l) + O(r^{*}_l - a)^2,
\end{eqnarray}

\noindent
and

\begin{eqnarray}
\nonumber
\Phi(r^{*}_l)
  &=& \Phi + 2\pi a B(a) (r^{*}_l - a)\\
\nonumber
  & &  + \pi(B(a) + aB'(a)) (r^{*}_l - a)^2 + O(r^{*}_l - a)^3\\
\label{Eq_A19}
  &=& (1 - \gamma) \Phi + O(1/N),
\end{eqnarray}

\noindent
and so $a - r^{*}_l = [\gamma\Phi / (\pi a|B'(a)|)]^{\half}$.
Substituting this result into (\ref{Eq_A18}) gives

\begin{equation}
\label{Eq_A20}
f''(r^{*}_l)
  = -2 e \left(\frac{\gamma\Phi|B'(a)|}{\pi a}\right)^{\half}
    +O(e\gamma).
\end{equation}

\noindent
Thus $f''(r^{*}_l) = O(e)$ for $\Lambda + 1 < l < (1 - \gamma)N$ and
so the first sum in (\ref{Eq_A17}) gives a contribution of $O(1)$.

Next consider the second term in (\ref{Eq_A17}). With $l = N-m-1$,

\begin{equation}
\label{Eq_A21}
\sum^{N-CN^{\frac{1}{3}}}_{l=(1-\gamma)N} \pdo{e} \ln |f''(r^{*}_l)|
  = \sum^{\gamma N}_{m = CN^{\frac{1}{3}}}  \pdo{e} \ln |f''(r^{*}_l)|.
\end{equation}

\noindent
The range of m in (\ref{Eq_A21}) is such that (\ref{Eq_A09}) is valid
so that

\begin{eqnarray}
\nonumber
\sum^{N-CN^{\frac{1}{3}}}_{l=(1-\gamma)N}\pdo{e} \ln |f''(r^{*}_l)|
  &=& \frac{\Phi}{4 \pi} \sum^{\gamma N}_{CN^{\frac{1}{3}}}
      \frac{1}{m} + O(1)\\
\label{Eq_A22}
  &=& \frac{\Phi}{6 \pi} \ln N + O(1).
\end{eqnarray}

\noindent
This completes the sum in (\ref{Eq_A17}) and the second sum in
(\ref{Eq_A13}).

Finally, consider the last sum in (\ref{Eq_A13}). This requires that
$I$ be estimated near the end point $l=N-1$ or $m=0$. For $N>>1$ and
with $\varphi(r)$ monotonically decreasing to zero
$(\varphi'(r) = -\Phi(r) / (2\pi r))$, the integral in (\ref{Eq_A01})
is dominated near $r = a$. Since $\varphi'(a) \neq 0$, $\varphi(r)$ has
a first-order zero at $r = a$: $\varphi(r) \sim (1-r/a)\Phi/2\pi$,
$r \to a$. Hence, for $N >> 1$

\begin{widetext}
\begin{equation}
\label{Eq_A23}
I(m=0) \sim 2^{-2N} (N + \epsilon)^{-2N} e^{2(N+\epsilon)}\\
  \int^{2(N+\epsilon)}_0 dx \, x^{2N-1} e^{-x}.
\end{equation}
\end{widetext}

\noindent
But,

\begin{equation}
\label{Eq_A24}
\int^{2(N+\epsilon)}_0 dx \, x^{2N-1} e^{-x} = (2N-1)!
  - \Gamma(2N,2(N+\epsilon)),
\end{equation}

\noindent
where $\Gamma(a,x)$ is the incomplete gamma function given by entry
6.5.3 in \cite{Abramowitz64}. Using entries 8.356.2 in
\cite{Gradshteyn65} and 6.5.35 in \cite{Abramowitz64},

\begin{equation}
\label{Eq_A25}
\Gamma(2N, 2(N+\epsilon))
  = e^{-2N} (2N)^{2N-1} (\sqrt{\pi N} + O(1)).
\end{equation}

\noindent
Combining (\ref{Eq_A23})-(\ref{Eq_A25}) with Stirling's formula gives

\begin{equation}
\label{Eq_A26}
I(m=0) \sim \half \sqrt{\frac{\pi}{N}} (1 + O(1/\sqrt{N})), N >> 1.
\end{equation}

\noindent
This is an overestimate as we integrated over all of the range $[0,a]$
instead of a patch near $r = a$, and therefore the factor
$\sqrt{\pi}/2$ in (\ref{Eq_A26}) cannot be trusted. However, the
result demonstrates that $I(m=0)$ falls off as a power of $N$ and not
exponentially. Since $I(m=0) < I(m=CN^{\frac{1}{3}})$ and
$I(m=CN^{\frac{1}{3}}) = O(N^{-\frac{1}{3}})$ we can state that
$\partial\ln I / \partial e = O(1/N)$ for
$0 \le m \le CN^{\frac{1}{3}}$ and so

\begin{equation}
\label{Eq_A27}
\sum^{CN^{\frac{1}{3}}}_{m=0} \pdo{e} \ln I = O(N^{-\frac{2}{3}}).
\end{equation}

Combining (\ref{Eq_A03}), (\ref{Eq_A04}), (\ref{Eq_A13}),
(\ref{Eq_A16}), (\ref{Eq_A17}), (\ref{Eq_A21}), (\ref{Eq_A22}),
(\ref{Eq_A27}) and intermediate results gives

\begin{equation}
\label{Eq_A28}
\sum^{N-1}_{l=0} \pdo{e} \ln I
  = 2e \int^{a}_0 dr \, r B(r) \varphi(r)
    - \frac{\Phi}{12\pi} \ln \left(\frac{e\Phi}{2\pi}\right)
    + O(\Lambda),
\end{equation}

\noindent
where $\Lambda >> 1$ but $e$-independant. This completes the sum of
the first term in (\ref{Eq_A02}).

The sum of the second term in (\ref{Eq_A02}) is straightforward:

\begin{eqnarray}
\nonumber
\sum^{N-1}_{l=0} \frac{1}{W-1}
  &=& \sum^{N-1}_0 \frac{1}{N + \epsilon - l-1}\\
\nonumber
  &=& \sum^{N-1}_{m=0} \frac{1}{m + \epsilon}\\
\label{Eq_A29}
  &=& \ln \left(\frac{e\Phi}{2\pi}\right) + O(1).
\end{eqnarray}

Now consider the sum of the third term in (\ref{Eq_A02}). Letting
$m = N - l - 1$,

\begin{widetext}
\begin{equation}
\label{Eq_A30}
\sum^{N-1}_{l=0}
  \left( \frac{e\Phi}{2\pi} - l - 1 + \frac{1}{2I} \right)^{-1}
  = \left( \sum^{CN^{\frac{1}{3}}}_{m=0}
  + \sum^{\gamma N}_{CN^{\frac{1}{3}}}
  + \sum^{N-1}_{\gamma N} \right) (m + \epsilon + g(m))^{-1},
\end{equation}
\end{widetext}

\noindent
where $1/g(m) = 2I$ and $g'(m) < 0$ for $0 \le m \le N-1$,
$\gamma << 1$, and $C$ is given by (\ref{Eq_A11}). Consider the first
sum:

\begin{eqnarray}
\nonumber
\lefteqn{\sum^{CN^{\frac{1}{3}}}_{m=0} (m + \epsilon + g(m))^{-1}}\\
\nonumber
  &<& \sum^{CN^{\frac{1}{3}}}_{m=0} (m + \epsilon + g(CN^{\frac{1}{3}}))^{-1}\\
\nonumber
  &<& \ln
  \left[
  \frac{CN^{\frac{1}{3}} + g(CN^{\frac{1}{3}}) + \epsilon}
       {g(CN^{\frac{1}{3}}) + \epsilon}
  \right] + O\left(\frac{1}{g(CN^{\frac{1}{3}})}\right)\\
\label{Eq_A31}
  &=& O(1),
\end{eqnarray} 

\noindent
since by definition of $g$ and (\ref{Eq_A10}),
$g(CN^{\frac{1}{3}}) = O(N^{\frac{1}{3}})$.

Next consider the second sum in (\ref{Eq_A30}). Since $g'(m) < 0$ we
have

\begin{eqnarray}
\nonumber
\lefteqn{\sum^{\gamma N}_{CN^{\frac{1}{3}}}
  (m + \epsilon + g(CN^{\frac{1}{3}}))^{-1}}\\
\nonumber
  &<& \sum^{\gamma N}_{CN^{\frac{1}{3}}} (m + \epsilon + g(m))^{-1}\\
\label{Eq_A32}
  &<& \sum^{\gamma N}_{CN^{\frac{1}{3}}} (m + \epsilon + g(\gamma N))^{-1}.
\end{eqnarray}

\noindent
The last sum is bounded by elementary means by noting that
$g(\gamma N) < g(CN^{\frac{1}{3}}) = O(N^{\frac{1}{3}})$ and hence
$g(\gamma N) / N^{\frac{1}{3}} = O(1)$ or less. Then by inspection the
right-hand side is bounded by $\frac{2}{3} \ln N + O(1)$. Likewise so
is the first sum, and hence

\begin{equation}
\label{Eq_A33}
\sum^{\gamma N}_{CN^{\frac{1}{3}}} (m + \epsilon + g(m))^{-1}
  = \frac{2}{3} \ln N + O(1).
\end{equation}

Finally, consider the last sum in (\ref{Eq_A30}). Again because
$g'(m) < 0$,

\begin{eqnarray}
\nonumber
\lefteqn{\sum^{N-1}_{\gamma N} (m + \epsilon + g(\gamma N))^{-1}}\\
\nonumber
  &<& \sum^{N-1}_{\gamma N} (m + \epsilon + g(m))^{-1}\\
\label{Eq_A34}
  &<& \sum^{N-1}_{\gamma N} (m + \epsilon + g(N-1))^{-1}.
\end{eqnarray}

\noindent
As $g(m) = 1 / (2I)$ and $CN^{\frac{1}{3}} < m = \gamma N << N$,
we can use (\ref{Eq_A10}) and conclude

\begin{widetext}
\begin{equation}
\label{Eq_A35}
g(\gamma N)
  = \left(\frac{\gamma a^3|B'(a)|}{4\pi\Phi}\right)^{\frac{1}{4}}
  N^{\half}
  \exp
  \left[
  - \frac{4}{3} \left(\frac{\Phi\gamma^3}{\pi a^3|B'(a)|}\right)^{\half}
  N + O\left(\gamma^{\half}\right)
  \right]
  \left[1 + O\left(\gamma^{\half}\right)\right].
\end{equation}
\end{widetext}

\noindent
Hence, $g(N-1) < g(\gamma N) = O(N^{\half} e^{-\lambda N})$ where
$\lambda = O(1)$. Simple estimates applied to the first and last sums
in (\ref{Eq_A34}) give

\begin{equation}
\label{Eq_A36}
\sum^{N-1}_{\gamma N} (m + \epsilon + g(m))^{-1} = - \ln \gamma + O(1/N).
\end{equation}

\noindent
Combining (\ref{Eq_A30}), (\ref{Eq_A31}), (\ref{Eq_A33}) and
(\ref{Eq_A36}) gives

\begin{equation}
\label{Eq_A37}
\sum^{N-1}_{l=0}
  \left( \frac{e\Phi}{2\pi} - l - 1 + \frac{1}{2I} \right)^{-1}
  = \frac{2}{3} \ln N + O(1).
\end{equation}

We now turn to the sum of the final term in (\ref{Eq_A02}). Using
previous definitions we can write this as

\begin{widetext}
\begin{equation}
\label{Eq_A38}
\sum^{N-1}_{l=0} \frac{1}{2I}
  \left( W - 1 - \frac{1}{2I} \right)^{-1} \pdo{e} \ln I
  =
  \left(
  \sum^{CN^{\frac{1}{3}}}_{m=0}
  + \sum^{\gamma N}_{CN^{\frac{1}{3}}}
  + \sum^{N-1}_{\gamma N}
  \right)
  \frac{g(m)}{m + \epsilon + g(m)} \pdo{e} \ln I.
\end{equation}
\end{widetext}

\noindent
Consider the first sum in (\ref{Eq_A38}). We have previously noted
that $\partial \ln I / \partial e = O(1/N)$ for the range of $m$
indicated. Since $0 < g(m) / (m + \epsilon + g(m)) \le 1$, then

\begin{equation}
\label{Eq_A39}
\sum^{CN^{\frac{1}{3}}}_{m=0}
  \frac{g(m)}{m + \epsilon + g(m)} \pdo{e} \ln I
  = O\left(N^{- \frac{2}{3}}\right).
\end{equation}

The range of $m$ in the second sum in (\ref{Eq_A38}) allows the use of
(\ref{Eq_A10}) for $I$, and hence

\begin{widetext}
\begin{equation}
\label{Eq_A40}
\pdo{e} \ln I
  = - \frac{\Phi}{8 \pi N} - \frac{\Phi}{8 \pi m}
  + \frac{\Phi}{24\pi C^{\frac{3}{2}}}
  \left[
  3 \sqrt{\frac{m}{N}} - \left(\frac{m}{N}\right)^{\frac{3}{2}}
  \right]
  + O\left(\frac{1}{\sqrt{mN}},\frac{\sqrt{m}}{N^{\frac{3}{2}}}\right),     
\end{equation}
\end{widetext}

\noindent
where $C$ is defined by (\ref{Eq_A11}). Then the second sum in
(\ref{Eq_A38}) is

\begin{widetext}
\begin{equation}
\label{Eq_A41}
\sum^{\gamma N}_{CN^{\frac{1}{3}}}
  \frac{g(m)}{m + \epsilon + g(m)} \pdo{e} \ln I
  = - \frac{\Phi}{8\pi} \sum^{\gamma N}_{CN^{\frac{1}{3}}}
  \left[
  \frac{1}{N} + \frac{1}{m} - C^{-\frac{3}{2}}
    \left(
    \sqrt{\frac{m}{N}} - \frac{1}{3}
    \left(\frac{m}{N}\right)^{\frac{3}{2}}
    \right)
  \right.
  \left.
  + O\left(\frac{1}{\sqrt{mN}},\frac{\sqrt{m}}{N^{\frac{3}{2}}}\right)
  \right]
  \frac{g(m)}{m + \epsilon + g(m)}.
\end{equation}
\end{widetext}

\noindent
For the range of $m$ indicated, $g(m) = 1 / (2I)$ is given by
(\ref{Eq_A10}) and has the functional form
$g(m) = \alpha (mN)^{\frac{1}{4}} \text{exp}(-\beta m^{\frac{3}{2}}/\sqrt{N})$, 
where $\alpha$, $\beta$ are constants. Note that
$(m + \epsilon + g(m))^{-1} < m^{-1}$. Then the first sum in
(\ref{Eq_A41}) vanishes as $N \to \infty$ by inspection. Over the
range of $m$ indicated, $1/m \le C^{-\frac{3}{2}} \sqrt{m/N}$ and
$(m/N)^{\frac{3}{2}} < (m/N)^{\half}$. Therefore the remaining sums in
(\ref{Eq_A41}) are dominated by
$\sum^{\gamma N}_{CN^{\frac{1}{3}}} g(m) / \sqrt{m N}$ which, when
approximated by an integral, is of $O(1)$ and so

\begin{equation}
\label{Eq_A42}
\sum^{\gamma N}_{CN^{\frac{1}{3}}}
  \frac{g(m)}{m + \epsilon + g(m)} \pdo{e} \ln I = O(1).
\end{equation}

Finally, we deal with the last sum in (\ref{Eq_A38}). It is for
$0 < l < (1 - \gamma) N$, and following (\ref{Eq_A03}) we estimated
$\partial\ln I / \partial e = O(1)$ or less for this $l$-range. We
have already noted that $g(\gamma N) = O(N^{\half} e^{-\lambda N})$
and that $g'(m) < 0$. Hence we conclude

\begin{eqnarray}
\nonumber
\sum^{N-1}_{\gamma N} \frac{g(m)}{m + \epsilon + g(m)}
\pdo{e} \ln I
  &\le& \sum^{N-1}_{\gamma N} \frac{g(m)}{m} \pdo{e} \ln I\\
\label{Eq_A43}
  &=& O( N^{\half} e^{-\lambda N}).
\end{eqnarray}

In summary, (\ref{Eq_A38}), (\ref{Eq_A39}), (\ref{Eq_A42}),
(\ref{Eq_A43}) give

\begin{equation}
\label{Eq_A44}
\sum^{N-1}_{l=0} \frac{1}{2I} \left( W - 1 + \frac{1}{2I} \right)^{-1}
  \pdo{e} \ln I = O(1).
\end{equation}

\noindent
Combining (\ref{Eq_A02}), (\ref{Eq_A28}), (\ref{Eq_A29}),
(\ref{Eq_A37}), (\ref{Eq_A44}) gives for $e\Phi/2\pi >> 1$

\begin{widetext}
\begin{equation}
\label{Eq_A45}
\sum^{N-1}_{l=0} \pdo{e} \ln ||\psi^0_l||^2
  = 2e \int^a_o dr \, r B(r) \varphi(r)
    - \frac{\Phi}{4 \pi} \ln \left(\frac{e\Phi}{2\pi}\right) + O(\Lambda),
\end{equation}
\end{widetext}

\noindent
where $\Lambda >> 1$ but $e$-independent. Now combine (\ref{Eq_A45})
with (\ref{Eq_0406}), integrate and combine this with our previous
result in (\ref{Eq_0407}) to get for $ma << 1$ followed by $e\Phi >> 1$

\begin{widetext}
\begin{equation}
\label{Eq_A46}
lndet = - \frac{e \Phi}{4 \pi} \ln \left(\frac{e\Phi}{2\pi}\right)
  + \frac{e\Phi}{4 \pi} \ln (ma)^2
  + O(e\Phi, (ma)^2 e\Phi \ln (e\Phi)).
\end{equation}
\end{widetext}

\noindent
The justification for the inclusion of the $\ln (ma)^2$ term was
discussed following (\ref{Eq_0407}). Also, as discussed immediately
after (\ref{Eq_0406}), there may be subtle cancellations that will
eliminate the $\ln (e\Phi)$ factor in the remainder term
$(ma)^2 e\Phi \ln (e\Phi)$. The case when $e\Phi < 0$ is included by
replacing $e \Phi$ in (\ref{Eq_A46}) everywhere with $|e\Phi|$.

This analysis is for fields $B(r) > 0$ for $r < a$ with continuous
first and second derivatives and with $B(a) = 0$, $B'(a) < 0$. For the
case $B(a) > 0$ the analysis is almost identical to the preceding case
and is also a little simpler. The main changes are

\begin{eqnarray}
\label{Eq_A47}
f(r^{*}) &=& \frac{m^2 \Phi}{2\pi N a^2 B(a)}
  + O\left(\frac{m^3}{N^2}\right),\\
|f''(r^{*})| &=& \frac{4 \pi N B(a)}{\Phi}
   \left(1 + O\left(\frac{m}{N}\right)\right)
\end{eqnarray}

\noindent
and

\begin{widetext}
\begin{equation}
\label{Eq_A49}
I = \left( \frac{\Phi}{2 N a^2 B(a)} \right)^{\half} \, \text{exp}
  \left[
  \frac{m^2 \Phi}{2\pi N a^2 B(a)} + O\left(\frac{m^3}{N^2}\right)
  \right]
  \left( 1 + O\left(\frac{m}{N}\right) \right),
\end{equation}
\end{widetext}

\noindent
provided $(a^2 B(a) /\Phi)^{\half} N^{\half} < m << N$. The result is
the same as (\ref{Eq_A45}).


\begin{thebibliography}{99}
\bibitem{Fry02}
M. P. Fry, Int. J. Mod. Phys. A \textbf{17}, 936 (2002).
\bibitem{Fry95}
M. P. Fry, Phys. Rev. D \textbf{51}, 810 (1995).
\bibitem{Chiu00}
T.W. Chiu, Nucl. Phys. \textbf{B588}, 400 (2000).
\bibitem{Sachs92}
I. Sachs and A. Wipf, Helv. Phys. Acta. \textbf{65}, 652 (1992).
\bibitem{Hausler01}
S. H\"{a}usler and C.B. Lang, Phys. Lett. B \textbf{515}, 213 (2001).
\bibitem{Fry92}
M. P. Fry, Phys. Rev. D \textbf{45}, 682 (1992); D \textbf{47}, 743(E) (1993).
\bibitem{Fry96}
M. P. Fry, Phys. Rev. D \textbf{54}, 6444 (1996).
\bibitem{Fry93}
M. P. Fry, Phys. Rev. D \textbf{47}, 2629 (1993).
\bibitem{Schwinger62}
J. Schwinger, Phys. Rev. \textbf{128}, 2425 (1962).
\bibitem{Musto86}
R. Musto, L. O'Raifeartaigh, and A. Wipf, Phys. Lett. B \textbf{175},
433 (1986).
\bibitem{Newton82}
R. G. Newton, \emph{Scattering Theory of Waves and Particles}, 2nd
ed. (Springer, New York, 1982).
\bibitem{Abramowitz64}
M. Abramowitz and I. A. Stegun, \emph{Handbook of Mathematical
Functions} (U.S. Government Printing Office, Washington, D.C., 1964).
\bibitem{Jaroszewicz86}
T. Jaroszewicz, Phys. Rev. D \textbf{34}, 3128 (1986).
\bibitem{Fry00a}
M. P. Fry, Phys. Rev. D \textbf{62}, 125007 (2000).
\bibitem{Aharonov79}
Y. Aharonov and A. Casher, Phys. Rev. A \textbf{19}, 2461 (1979).
\bibitem{Seiler75}
E. Seiler and B. Simon, Commun. Math. Phys. \textbf{45}, 99 (1975).
\bibitem{Seiler82}
E. Seiler, \emph{Gauge Theories as a Problem of Constructive Quantum Field
Theory and Statistical Mechanics}, Lecture Notes in Physics Vol. 159
(Springer, Berlin, 1982).
\bibitem{Seiler81}
E. Seiler, in \emph{Gauge Theories: Fundamental Interactions and Rigorous
Results}, Proceedings of the International Summer School of Theoretical
Physics, Poiana Brasov, Romania, 1981, edited by P. Dita,
V. Georgescu, and R. Purice, Progress in Physics Vol. 5 (Birkh\"{a}user,
Boston, 1982), p. 263.
\bibitem{Brydges79}
D. Brydges, J. Fr\"{o}hlich, and E. Seiler, Ann. Phys. (N.Y.)
\textbf{121}, 227 (1979).
\bibitem{Weingarten80}
D. Weingarten, Ann. Phys. (N.Y.) \textbf{126}, 154 (1980).
\bibitem{Gradshteyn65}
I. S. Gradshteyn and I. M. Ryzhik, \emph{Table of Integrals, Series
and Products} (Academic Press, New York, 1965).
\bibitem{Prudnikov88}
A. P. Prudnikov, Yu. A. Brychkov, and O. I. Marichev, \emph{Integrals
and Series} (Gordon and Breach, New York, 1988), Vol. 2.
\bibitem{Fry00b}
M. P. Fry, J. Math. Phys. \textbf{41}, 1691 (2000).
\bibitem{Erdos93}
L. Erd\"{o}s, Lett. Math. Phys. \textbf{29}, 219 (1993).
\bibitem{Schwinger51}
J. Schwinger, Phys. Rev. \textbf{82}, 664 (1951).
\bibitem{Jayewardena88}
C. Jayewardena, Helv. Phys. Acta. \textbf{61}, 636 (1988).
\bibitem{Montonen77}
C. Montonen and D. Olive, Phys. Lett. B \textbf{72}, 117 (1977);
P. Goddard, J. Nuyts, and D. Olive, Nucl. Phys. \textbf{B125}, 1
(1977).
\bibitem{Dunne98}
G. Dunne and T. Hall, Phys. Rev. D \textbf{58}, 105022 (1998).
\bibitem{Dunne99}
G. Dunne and T. Hall, Phys. Rev. D \textbf{60}, 065002 (1999).
\bibitem{Martin88}
C. Martin and D. Vautherin, Phys. Rev. D \textbf{38}, 3593 (1988).
\end{thebibliography}
\end{document}